\documentclass[authoryear,review,3p,times]{elsarticle}

\usepackage{amssymb}
\usepackage{amsmath}
\usepackage{lineno}
\usepackage{bm} 
\usepackage{longtable}
\usepackage{mathtools}
\usepackage{xfrac}
\usepackage{multirow}
\usepackage{caption}
\usepackage{soul}
\usepackage{color}
\usepackage{relsize}
\usepackage{mathtools}  
\usepackage{makecell}
\usepackage{placeins}

\usepackage{algorithm}
\usepackage{algpseudocode}

\usepackage{array}
\usepackage{booktabs}

\newcommand{\E}[1]{\text{E}\left[ #1 \right]}
\newcommand{\Var}[1]{\text{Var}\left[ #1 \right]}

\newcommand{\corr}[1]{\text{corr}\left[ #1 \right]}

\newcommand{\mat}[1]{{#1}}
\renewcommand{\vec}[1]{\bm{#1}}

\newcommand{\safeincludegraphics}[2][]{%
  \IfFileExists{#2}{\includegraphics[#1]{#2}}{\fbox{\parbox{0.85\linewidth}{\centering Missing figure file: \texttt{\detokenize{#2}}}}}%
}

\journal{Annals of Nuclear Energy}

\begin{document}

\begin{frontmatter}



\title{Second-Order Correction for Statistical Bias in Monte Carlo Depletion Calculations}


\author[MIT,UGD]{Esthem Bouvet} 
\author[MIT]{Benoit Forget}
\author[MIT]{Dean Price\corref{cor1}}
\cortext[cor1]{Corresponding author}
\ead{dnprice@mit.edu}

\affiliation[MIT]{organization={Department of Nuclear Science and Engineering, Massachusetts Institute of Technology},
            addressline={60 Vassar St}, 
            city={Cambridge},
            postcode={02139}, 
            state={MA},
            country={US}}
\affiliation[UGD]{organization={Institut National des Sciences et Techniques Nucléaires (INSTN)},
            addressline={Centre CEA de Saclay},
            city={Gif-sur-Yvette},
            postcode={91191},
            country={France}}

\begin{abstract}
Depletion analyses are often driven by Monte Carlo neutron transport methods that introduce uncertainty into the transmutation matrix used to evolve a material composition. 
The nonlinearity of the matrix exponential applied to this transmutation matrix causes the uncertainty to introduce a systematic bias in the computed isotope vectors.
This work derives a second-order correction for this bias that can be computed with quantities obtainable from typical Monte Carlo transport routines.
An augmented Bateman system is formulated to evaluate these quantities and an adapted Chebyshev rational approximation method uses transmutation matrix factorizations for efficient evaluation of the second-order correction.
Then, the correction is demonstrated using a ten-nuclide verification problem, an ITER port-plug activation problem, and a 9-pin boiling water reactor lattice containing self-shielded gadolinium-bearing fuel.
\end{abstract}


\begin{highlights}
\item Monte Carlo reaction-rate noise systematically biases isotope inventories.
\item A second-order correction accounts for correlated reaction-rate uncertainty.
\item An adapted Chebyshev rational approximation reuses matrix factorizations.
\item Realistic applications demonstrate substantial reductions in global inventory bias.
\end{highlights}


\end{frontmatter}




\section{Introduction}
\label{sec:intro}
The matrix exponential solution to the Bateman equations provides the mathematical foundation for nuclide transmutation calculations.
After reaction rates are determined from a neutron transport calculation, this formulation is used to evolve nuclide inventories in applications such as activation analysis \citep{pettinari2024neutron, quang2023activation}, fission reactor depletion \citep{hoffman2026improving, price2024multiphysics}, spent fuel analysis \citep{radaideh2019advanced, price2019advanced}, and medical isotope production \citep{reichenberger2022canal, labossierehickman2024mccafe}.
Many of these workflows rely on stochastic methods to estimate reaction rates, which introduces unavoidable uncertainty into the inputs to the Bateman equations.
As the matrix exponential is nonlinear, conventional approaches that construct transmutation behavior from mean reaction rates introduce bias into the final nuclide inventories.
This work presents a higher-order correction that reduces this bias and may be useful for problems with large reaction rate uncertainties in regions of interest.

The interaction between the uncertainty in Monte Carlo based transport solution methods and the Bateman equations has been studied before.
\citet{xu2010statistical} provides an overview of this challenge and performs computational experiments to gain an empirical understanding of how the number density uncertainties behave across multiple transport/Bateman iterative timesteps.
Notably, Xu et al. showed that biases in nuclide concentrations can accumulate over successive depletion steps (see Table II in this reference).
In \citet{cosgrove2020neutron}, the stochasticity created by these uncertainties is discussed in the context of their ability to create instabilities in depletion calculations.
A review of studies concerning general instabilities arising in transport/Bateman coupling is provided in \cite{du2023monte} alongside some test cases to explore these instabilities for a pressurized water reactor.
Some studies are also concerned with quantifying this uncertainty as it propagates through transport-Bateman iterations.
An approach to this propagation that is limited to a particular set of isotopes is presented in \citet{tohjoh2006effect}.
Comprehensive first order approaches have also been explored to directly propagate this uncertainty through the Bateman equations \citet{park2018monte, garcia2008propagation, takeda1999estimation}.
Although much of the discussion present in the current work can enable the propagation of this uncertainty, uncertainty propagation is not the focus.

Instead, this work focuses on the bias introduced when the matrix exponential is evaluated using a transmutation matrix constructed from mean reaction rates.
This neglects second- and higher-order contributions of reaction rate uncertainty to the expected nuclide densities.
Notably, this evaluation approach is unbiased to first order, so the resulting bias is generally small.
However, its effect may be measurable in depletable zones with large reaction-rate uncertainties and accumulates \citep{xu2010statistical} over many depletion steps.
Additionally, the present study evaluates the correction only for single constant-matrix depletion steps.
Although many higher-order depletion schemes may use depletion matrices calculated at multiple previous steps to advance the nuclide concentrations, they are not considered in the current study.
Further mathematical discussion on this bias is provided in \citet{dumonteil2011biases} where a correction based on a scalar flux perturbation is presented.
Expanding on this previous work, the present formulation expresses the stochasticity in the transmutation matrix with multiple reaction rate estimators that span different nuclides and reaction channels.
Then, a multivariate expansion of the matrix exponential with respect to these correlated reaction rate estimators is used to form a second-order correction to the expected nuclide densities.
The objective of the current work is to derive this correction and identify the quantities required for its calculation.
The computational burden of the resulting augmented Bateman system is reduced with an adapted CRAM algorithm that reuses transmutation-matrix factorizations across the additional equation systems the correction introduces to the typical Bateman equations.
Three demonstration problems are provided to illustrate the application of the method.
First, a ten-nuclide toy problem is presented with all information required to verify independent implementations.
The second demonstration application presents a material activation in an ITER port plug.
Finally, the correction is evaluated for a 9-pin boiling water reactor (BWR) lattice containing a self-shielded gadolinium-bearing fuel pin.
In this eigenvalue demonstration, intergenerational correlations are avoided with a brute force method.
These correlations may be problematic if left untreated in more conventional workflows.

The remainder of the paper is organized as follows.
Section~\ref{sec:method} presents the stochastic Bateman formulation and the augmented system used to compute the correction.
Section~\ref{sec:demonstration} applies the method to three demonstration problems.
Section~\ref{sec:conc} summarizes the main conclusions and outlines future work.

\section{Methodology}
\label{sec:method}
\subsection{Matrix Exponential and Bateman Equations}
\label{sec:matrex_bateman}
The Bateman equations form a system of coupled first order ordinary differential equations that can be used to describe transmutations within a population of isotopes.
For a period with fixed irradiation conditions, these equations are expressed with
\begin{gather}
\label{eq:bateman}
\frac{d \vec{n}(t)}{dt} = \mat{Q}\,\vec{n}(t)
\end{gather}
where $\vec{n} \in \mathbb{R}^K$ is used to represent a vector of $K$ isotopic concentrations and $\mat{Q} \in \mathbb{R}^{K \times K}$ contains the transmutation rate coefficients that govern the production and removal of isotopes.
The initial conditions to this system of equations are provided with
\begin{gather}
\vec{n}(t=0) = \vec{n}_0 
\end{gather}
where $\vec{n}_0 \in \mathbb{R}^K$.

For this discussion, it is convenient to decompose the transmutation matrix into a portion that collects all contributions from nuclear decay and a portion that collects all contributions from neutron-induced reactions with
\begin{gather}
\mat{Q} = \mat{\Lambda} + \mat{\Gamma}.
\label{eq:Qsplit}
\end{gather}
The matrix $\mat{\Lambda} \in \mathbb{R}^{K \times K}$ contains only decay information and can be further specified with
\begin{gather}
\left(\mat{\Lambda}\right)_{i,i'} =
\begin{cases}
\lambda_{i' \rightarrow i}, & i \ne i' \\
-\mathlarger{\sum_{k = 1, k \ne i}^K} \lambda_{i \rightarrow k}, & i = i'
\end{cases}.
\label{eq:Lambda_def}
\end{gather}
Here, $\left(\mat{\Lambda}\right)_{i,i'}$ is used to indicate the entry within $\mat{\Lambda}$ located in the $i$-th row and $i'$-th column.
Then, $\lambda_{i' \rightarrow i}$ is used to represent the decay constant of the $i'$-th isotope decaying into the $i$-th isotope.
The matrix $\mat{\Gamma} \in \mathbb{R}^{K \times K}$ can be specified similarly
\begin{gather}
\left(\mat{\Gamma}\right)_{i,i'} =
\begin{cases}
r_{i' \rightarrow i}, & i \ne i', \\
-r_{i'}^{\mathrm{rem}}, & i = i'.
\end{cases}
\label{eq:Gamma_def}
\end{gather}
Here, $r_{i' \rightarrow i}$ is used to represent the neutron-induced production rate for transmutation from the $i'$-th isotope to the $i$-th isotope and $r_{i'}^{\mathrm{rem}}$ is the total neutron-induced removal rate coefficient of nuclide $i'$.

With $Q$ fully described, the solution to Equation \ref{eq:bateman} can be expressed using the matrix exponential
\begin{gather}
\vec{n}(t) = \exp\left(\mat{Q} t \right)\vec{n}_0.
\end{gather}

\subsection{Transmutation Matrix Construction}
Practically, when using Monte Carlo codes within transmutation analyses, an exact value of $\mat{\Gamma}$ is inaccessible.
An estimate is formed from the finite number of sampled histories tracked during the particle transport routine.
In OpenMC \citep{romano2015openmc}, this matrix is formed by tallying energy-integrated transmutation reactions within each material zone and applying constant representative branching ratios \citep{romano2021depletion} for isomeric states.
A similar approach is used in Serpent 2 \citep{leppanen2025status} but with the ability to include energy dependent isomeric branching ratios with the \texttt{set bralib} card.
The energy dependent isomeric branching is also available in MCNP \citep{fensin2011new}.
For fission-induced transmutations, consideration of the incident neutron is more commonly accounted for when selecting fission product yields.
\citet{romano2021depletion} presents some wider discussion on this topic.
For example, MCNP \citep{hendricks2008mcnpx} uses representative fission product yields based on a thermal, fast and high energy categorization determined by which energy range contains the majority of fission reactions.

Regardless of the method used, $\mat{\Gamma}$ can be expressed as a linear summation of tallies.
For brevity, the discussion in the current work will be presented without consideration for energy-dependent branching ratios or fission product yields but it is a trivial mathematical extension.
With this specification,
\begin{gather}
\label{eq:gammahatexpr}
\mat{\Gamma}(a) = \sum_{i'=1}^K \sum_x \vec{c}_{i'}^{(\mathrm{n},x)}\vec{e}_{i'}^T a_{i'}^{(\mathrm{n},x)}
\end{gather}
where
\begin{gather}
\vec{c}_{i'}^{(\mathrm{n},x)} \equiv \vec{b}_{i'}^{(\mathrm{n},x)}-\vec{e}_{i'}.
\end{gather}
Here, $a_{i'}^{(\mathrm{n},x)}$ denotes the $(\mathrm{n},x)$ per-parent-nuclide reaction rate for nuclide $i'$ and $a \equiv \left\{a_i^{(\mathrm{n}, x)}\right\}$ represents an arbitrary collection of reaction rates across every isotope and reaction type.
Then, $\vec{b}_{i'}^{(\mathrm{n},x)} \in \mathbb{R}^K$ is the product yield of this reaction that represents the expected number of nuclides of a certain type produced per reaction and $\vec{e}_{i} \in \mathbb{R}^K$ is used to represent the $i$-th standard basis vector and $\sum_x$ denotes a summation over all $N_{\mathrm{x}}$ total reaction channels.
It is also useful to combine Equation \ref{eq:gammahatexpr} with $\Lambda$ to express a transmutation matrix, formed from a particular set of reaction rates as
\begin{gather}
Q(a) = \Lambda + \sum_{i'=1}^K \sum_x \vec{c}_{i'}^{(\mathrm{n},x)}\vec{e}_{i'}^T a_{i'}^{(\mathrm{n},x)}.
\end{gather}

\subsection{Reaction-Rate Quantities}
\label{sec:rrq}
In this work, three reaction rate collections can be defined.
\begin{enumerate}
\item $r \equiv \left\{r_i^{(\mathrm{n}, x)}\right\}$ describes the reaction rates for the exact transport solution. These reaction rates are not directly available from the Monte Carlo transport calculation due to the finite number of sampled particle histories.
\item $\hat{R} \equiv \left\{\hat{R}_i^{(\mathrm{n}, x)}\right\}$ describes a collection of random estimators corresponding to the stochastic reaction rates produced from the Monte Carlo transport calculation.
\item $\hat{r}\equiv\left\{\hat{r}_i^{(\mathrm{n}, x)}\right\}$ describes a specific set of realizations of $\hat{R}$. This $\hat{r}$ would be the numerical result from a Monte Carlo calculation.
\end{enumerate}
As mentioned previously, the estimators $\hat{R}_{i'}^{(\mathrm{n},x)}$ are unbiased such that, 
\begin{gather}
\E{\hat{R}_{i'}^{(\mathrm{n},x)}} = r_{i'}^{(\mathrm{n},x)}.
\end{gather}

For the methods presented in the following sections, it is useful to provide some discussion on the second-order moments of $\hat{R}_{i'}^{(\mathrm{n},x)}$.
In general, the estimators $\hat{R}_{i'}^{(\mathrm{n},x)}$ are not independent.
The exact dependence structure may depend on the tallying approach, but, theoretically,
\begin{gather}
\left(s_{i}^{(\mathrm{n},x)}\right)^2 \equiv \Var{\hat{R}_{i}^{(\mathrm{n},x)}}
\end{gather}
and
\begin{gather}
\rho_{ix,jy} \equiv \corr{\hat{R}_{i}^{(\mathrm{n},x)}, \hat{R}_{j}^{(\mathrm{n},y)}}
\end{gather}
can be defined.
Sample estimates of these quantities, denoted by $\left(\hat{s}_{i}^{(\mathrm{n},x)}\right)^2$ and $\hat{\rho}_{ix,jy}$, respectively, can be formed using generations or batches of the particle transport calculation.
However, in eigenvalue calculations, generations of the Monte Carlo transport calculation are not strictly independent.
Therefore, treating generation-wise tallies as independent can bias the resulting variance estimates \citep{brissenden1986biases}, as introduced in Section \ref{sec:BWR-lattice}.
The correction derived in this work assumes that estimates of these statistical inputs are available and does not seek to improve the methods associated with their estimation.

A solution to the Bateman system,
\begin{gather}
\label{eq:batemanp}
\frac{d\bm{n}(t; a)}{dt}=Q(a)\bm{n}(t;a) \\
\bm{n}(t=0; a) = \bm{n}_0, \label{eq:batemanpic}
\end{gather}
that is dependent on a specific set of reaction rates, $a$,  can be represented with $\bm{n}(t; a)$ and can be expressed as
\begin{gather}
\bm{n}(t;a) = \exp\left(Q(a) t \right)\bm{n}_0.
\end{gather}

\subsection{Bias Correction}
\label{sec:bias}
Conventional approaches use $\bm{n}(t;\hat{r})$ without consideration of any bias introduced by nonlinearities of second order and higher in the matrix exponential.
To derive the higher order correction, use a series expansion of $\bm{n}(t; \hat{R})$ about $r$,
\begin{gather}
\label{eq:Nexpand}
\bm{n}(t; \hat{R}) \approx \vec{n}(t; r) + \sum_{i'=1}^K \sum_x \left( \hat{R}_{i'}^{(\mathrm{n},x)} - r_{i'}^{(\mathrm{n},x)}\right) \vec{u}_{i'}^{(\mathrm{n},x)}(t; r) + \frac{1}{2}\sum_{i'=1}^K \sum_x \sum_{j'=1}^K \sum_y \left( \hat{R}_{i'}^{(\mathrm{n},x)} - r_{i'}^{(\mathrm{n},x)}\right)\left( \hat{R}_{j'}^{(\mathrm{n},y)} - r_{j'}^{(\mathrm{n},y)}\right) \vec{v}_{i',j'}^{(\mathrm{n},x),(\mathrm{n},y)}(t;r)
\end{gather}
where
\begin{gather}
\vec{u}_i^{(\mathrm{n},x)}(t;a) \equiv \frac{\partial \bm{n}(t;a)} {\partial a_i^{(\mathrm{n},x)}}
\end{gather}
and
\begin{gather}
\vec{v}_{i,j}^{(\mathrm{n},x),(\mathrm{n},y)}(t;a) \equiv \frac{\partial^2 \bm{n}(t;a)} {\partial a_i^{(\mathrm{n},x)}  \partial a_j^{(\mathrm{n},y)}}.
\end{gather}
Now, applying the expectation operator to Equation \ref{eq:Nexpand} makes the leading second-order bias apparent.
A computable approximation to the corresponding correction is obtained by evaluating the sensitivities at the realized reaction rates $\hat{r}$ and replacing the population statistics $s_i^{(\mathrm{n},x)}$ and $\rho_{ix,jy}$ with their estimates $\hat{s}_i^{(\mathrm{n},x)}$ and $\hat{\rho}_{ix,jy}$,
\begin{gather}
\label{eq:correct}
\delta \bm{n}^{(2)}(t;\hat{r}) \equiv -\frac{1}{2}\sum_{i'=1}^K \sum_x \sum_{j'=1}^K \sum_y \hat{\rho}_{i'x,j'y}\hat{s}_{i'}^{(\mathrm{n},x)}\hat{s}_{j'}^{(\mathrm{n},y)}\vec{v}_{i',j'}^{(\mathrm{n},x),(\mathrm{n},y)}(t;\hat{r}).
\end{gather}
A second-order bias-reduced estimate of the isotope concentrations is therefore
\begin{gather}
\label{eq:corrected}
\bm{n}^{(2)}(t;\hat{r}) \equiv \bm{n}^{(0)}(t;\hat{r}) + \delta \bm{n}^{(2)}(t;\hat{r}),
\end{gather}
where $\bm{n}^{(0)}(t;\hat{r}) \equiv \bm{n}(t;\hat{r})$ denotes the conventional zeroth-order unbiased Bateman solution evaluated using the realized reaction rates.

\subsection{Augmented Bateman System}
Differentiating Equation \ref{eq:batemanp} with respect to $a_i^{(\mathrm{n},x)}$ and evaluating at $a=\hat{r}$ to yield a governing equation with associated initial conditions for each $\vec{u}_i^{(\mathrm{n},x)}(t;\hat{r})$,
\begin{gather}
\frac{d\vec{u}_i^{(\mathrm{n},x)}(t;\hat{r})}{dt} = Q(\hat{r})\vec{u}_i^{(\mathrm{n},x)}(t;\hat{r}) + \vec{c}_i^{(\mathrm{n},x)}\vec{e}_i^T\bm{n}^{(0)}(t;\hat{r}), \label{eq:ug}\\
\vec{u}_i^{(\mathrm{n},x)}(0;\hat{r}) = \vec{0}. \label{eq:uic}
\end{gather}
Next, differentiating the first-order sensitivity equation with respect to $a_j^{(\mathrm{n},y)}$ and evaluating at $a=\hat{r}$ yields a governing equation with associated initial conditions for each $\vec{v}_{i,j}^{(\mathrm{n},x),(\mathrm{n},y)}(t;\hat{r})$,
\begin{gather}
\label{eq:2ord}
\frac{d\vec{v}_{i,j}^{(\mathrm{n},x),(\mathrm{n},y)}(t;\hat{r})}{dt} = Q(\hat{r})\vec{v}_{i,j}^{(\mathrm{n},x),(\mathrm{n},y)}(t;\hat{r}) + \vec{c}_i^{(\mathrm{n},x)}\vec{e}_i^T\vec{u}_j^{(\mathrm{n},y)}(t;\hat{r}) + \vec{c}_j^{(\mathrm{n},y)}\vec{e}_j^T\vec{u}_i^{(\mathrm{n},x)}(t;\hat{r}),\\
\vec{v}_{i,j}^{(\mathrm{n},x),(\mathrm{n},y)}(0;\hat{r}) = \vec{0}.
\end{gather}
It is not necessary to evaluate each $\vec{v}_{i',j'}^{(\mathrm{n},x),(\mathrm{n},y)}(t;\hat{r})$.
Instead, it is more efficient to take the time derivative of Equation \ref{eq:correct} and substitute $\frac{d\vec{v}_{i,j}^{(\mathrm{n},x),(\mathrm{n},y)}(t;\hat{r})}{dt}$ from Equation \ref{eq:2ord}. Using the symmetry of the estimated covariance terms allows the two sensitivity contributions to be combined, forming a single first-order differential equation for $\delta\bm{n}^{(2)}(t;\hat{r})$,
\begin{gather}
\frac{d\delta\bm{n}^{(2)}(t;\hat{r})}{dt} = Q(\hat{r})\delta\bm{n}^{(2)}(t;\hat{r}) - \sum_{i'=1}^K\sum_x\sum_{j'=1}^K\sum_y\hat{\rho}_{i'x,j'y}\hat{s}_{i'}^{(\mathrm{n},x)}\hat{s}_{j'}^{(\mathrm{n},y)}\vec{c}_{i'}^{(\mathrm{n},x)}\vec{e}_{i'}^T\vec{u}_{j'}^{(\mathrm{n},y)}(t;\hat{r}), \label{eq:cog}\\
\delta\bm{n}^{(2)}(0;\hat{r}) = \vec{0}. \label{eq:cig}
\end{gather}

Equations \ref{eq:batemanp}, \ref{eq:ug}, and \ref{eq:cog} form the complete system of equations required to evaluate $\delta\bm{n}^{(2)}(t;\hat{r})$, with initial conditions provided in Equations \ref{eq:batemanpic}, \ref{eq:uic}, and \ref{eq:cig}.
With block notation, these equations can be expressed as
\begin{gather}
\frac{d\vec{z}(t;\hat{r})}{dt} = \mat{A}(\hat{r})\vec{z}(t;\hat{r}), \label{eq:augmentgov}\\
\vec{z}_0 \equiv \vec{z}(0;\hat{r}) = \begin{bmatrix} \bm{n}_0 & \left[\vec{0}\right]^T_{\substack{i=1,\ldots,K\\x}} & \vec{0} \end{bmatrix}^T, \label{eq:augmentic}
\end{gather}
where
\begin{gather}
\vec{z}(t;\hat{r}) \equiv \begin{bmatrix} \bm{n}^{(0)}(t;\hat{r}) & \left[\vec{u}_i^{(\mathrm{n},x)}(t;\hat{r})\right]^T_{\substack{i=1,\ldots,K\\x}} & \delta\bm{n}^{(2)}(t;\hat{r}) \end{bmatrix}^T
\end{gather}
and
\begin{gather}
\label{eq:adef}
\mat{A}(\hat{r}) \equiv \begin{bmatrix}
Q(\hat{r}) & \left[\mat{0}\right]_{\substack{i=1,\ldots,K\\x}} & \mat{0} \\
\left[\vec{c}_i^{(\mathrm{n},x)}\vec{e}_i^T\right]_{\substack{i=1,\ldots,K\\x}} & \left[\begin{cases}Q(\hat{r}) & i=j\text{ and }x=y \\ \mat{0} & \text{else}\end{cases}\right]_{\substack{i=1,\ldots,K\\j=1,\ldots,K\\x\\y}} & \mat{0} \\
\mat{0} & \left[-\sum_{i'=1}^K\sum_x\hat{\rho}_{i'x,jy}\hat{s}_{i'}^{(\mathrm{n},x)}\hat{s}_j^{(\mathrm{n},y)}\vec{c}_{i'}^{(\mathrm{n},x)}\vec{e}_{i'}^T\right]_{\substack{j=1,\ldots,K\\y}} & Q(\hat{r})
\end{bmatrix}.
\end{gather}
The solution can then be expressed as
\begin{gather}
\vec{z}(t;\hat{r}) = \exp\left(\mat{A}(\hat{r})t\right)\vec{z}_0.
\end{gather}
The notation $\left[\vec{u}_i^{(\mathrm{n},x)}(t;\hat{r})\right]_{\substack{i=1,\ldots,K\\x}}$ is used to denote the vertical concatenation of all $\vec{u}_i^{(\mathrm{n},x)}(t;\hat{r})$, and
\begin{gather}
\left[\begin{cases}Q(\hat{r}) & i=j\text{ and }x=y \\ \mat{0} & \text{else}\end{cases}\right]_{\substack{i=1,\ldots,K\\j=1,\ldots,K\\x\\y}}
\end{gather}
is used to denote the $K^2N_{\mathrm{x}}\times K^2N_{\mathrm{x}}$ block diagonal matrix with $Q(\hat{r})$ occupying the diagonal blocks.
In these equations, $\vec{z}(t;\hat{r})\in\mathbb{R}^{K(2+KN_{\mathrm{x}})}$ is used to track $\bm{n}^{(0)}(t;\hat{r})$, each $\vec{u}_i^{(\mathrm{n},x)}(t;\hat{r})$, and $\delta\bm{n}^{(2)}(t;\hat{r})$.
Once $\vec{z}(t;\hat{r})$ is evaluated, $\bm{n}^{(0)}(t;\hat{r})$ and $\delta\bm{n}^{(2)}(t;\hat{r})$ can be extracted and used in Equation \ref{eq:corrected} to recover the bias-corrected estimate.
For ease of implementation, a less compact presentation of Equation \ref{eq:adef} is provided in \ref{ap:betterpres}.
Then, $\mat{A}(\hat{r})\in\mathbb{R}^{K(2+KN_{\mathrm{x}})\times K(2+KN_{\mathrm{x}})}$ is a block lower triangular matrix whose diagonal blocks are repeated copies of $Q(\hat{r})$.
This leads to a notable relationship between $\mat{A}(\hat{r})$ and $Q(\hat{r})$ in that they share eigenvalues that differ in their multiplicities by a factor of $KN_{\mathrm{x}}+2$.
As the distribution of the eigenvalues of $Q(\hat{r})$ motivates the use of the Chebyshev rational approximation method (CRAM) to compute $\exp\left(Q(\hat{r})t\right)\bm{n}_0$ \citep{pusa2013numerical} in many modern codes, it is natural to use CRAM to also treat $\exp\left(\mat{A}(\hat{r})t\right)\vec{z}_0$.
This is the subject of the next section.

\subsection{Computational Treatment of Augmented Bateman System}
\label{sec:comptreat}
This section will provide some guidance on the evaluation of $\exp\left(\mat{A}(\hat{r})t\right)\vec{z}_0$ to alleviate some of the computational burden associated with this operation.
Several implementation efficiencies are not discussed in detail.
For example, $\mat{A}(\hat{r})$ should be represented using a sparse matrix and the summations over isotope $i$ and reaction $x$ can be restricted to active reaction channels and isotopes in the depletion system.
However, modifications to the typical CRAM algorithm can yield an efficiency due to the unique structure of $\mat{A}(\hat{r})$.

Specifically, the block lower triangular structure of $\mat{A}(\hat{r})$ avoids the need to perform linear solves on a full $K(2+KN_{\mathrm{x}})\times K(2+KN_{\mathrm{x}})$ system and instead allows for $(2+KN_{\mathrm{x}})$ solution routines, each performed on a $K\times K$ system.
Another efficiency gain is available because these additional linear systems can be expressed as shifted versions of those associated with calculating $\bm{n}^{(0)}(t;\hat{r})$ and therefore be accelerated using a precomputed factorization of $tQ(\hat{r})-\theta_\ell\mat{I}_K$ for each of the $\theta_\ell$ CRAM poles.
An adapted version of the incomplete partial fraction form of CRAM \citep{maria2016higher, romano2021depletion} that enables these efficiencies is given in Algorithm \ref{alg:newcram}.
For comparison, the unmodified version of the CRAM algorithm is presented in Algorithm \ref{alg:oldcram}.
Both the $2L$-th order adapted and unmodified CRAM algorithms rely on $L$ tabulated poles and incomplete-partial-fraction coefficients, $\theta_\ell,\widetilde{\alpha}_\ell\in\mathbb{C}$, together with $\alpha_0\in\mathbb{R}$, as provided in \citet{maria2016higher}.
More discussion on this approach is provided in \ref{ap:aipfcram}.
For compactness, the explicit dependence of the solution quantities on $t$ and $\hat{r}$ is suppressed within the algorithms.

\begin{algorithm}
\caption{Description of adapted $2L$-th order CRAM for treatment of the augmented Bateman system.}
\label{alg:newcram}
\begin{algorithmic}[1]
\State $\bm{n}^{(0)} \gets \bm{n}_0$
\State $\vec{u}_{i}^{(\mathrm{n},x)} \gets \vec{0}$ for each active $(i,x)$
\State $\delta\bm{n}^{(2)} \gets \vec{0}$
\For{$\ell \gets 1$ to $L$}
    \State Form and factorize $\mat{S}_{\ell} \gets tQ(\hat{r})-\theta_{\ell}\mat{I}_K$
    \State Solve $\mat{S}_{\ell}\vec{y}_{\ell}^{(0)}=\bm{n}^{(0)}$ for $\vec{y}_{\ell}^{(0)}$
    \For{each active $(i,x)$}
        \State Solve $\mat{S}_{\ell}\vec{y}_{i,\ell}^{(\mathrm{n},x)}=\vec{u}_{i}^{(\mathrm{n},x)}-t\vec{c}_i^{(\mathrm{n},x)}\vec{e}_i^T\vec{y}_{\ell}^{(0)}$ for $\vec{y}_{i,\ell}^{(\mathrm{n},x)}$
    \EndFor
    \State Solve $\displaystyle \mat{S}_{\ell}\vec{y}_{\ell}^{(2)}=\delta\bm{n}^{(2)}+t\sum_{\substack{(j,y)\,\mathrm{active}}}\sum_{\substack{(i,x)\,\mathrm{active}}}\hat{\rho}_{ix,jy}\hat{s}_i^{(\mathrm{n},x)}\hat{s}_j^{(\mathrm{n},y)}\vec{c}_i^{(\mathrm{n},x)}\vec{e}_i^T\vec{y}_{j,\ell}^{(\mathrm{n},y)}$ for $\vec{y}_{\ell}^{(2)}$
    \State $\bm{n}^{(0)} \gets \bm{n}^{(0)}+2\operatorname{Re}\!\left(\widetilde{\alpha}_{\ell}\vec{y}_{\ell}^{(0)}\right)$
    \For{each active $(i,x)$}
        \State $\vec{u}_{i}^{(\mathrm{n},x)} \gets \vec{u}_{i}^{(\mathrm{n},x)}+2\operatorname{Re}\!\left(\widetilde{\alpha}_{\ell}\vec{y}_{i,\ell}^{(\mathrm{n},x)}\right)$
    \EndFor
    \State $\delta\bm{n}^{(2)} \gets \delta\bm{n}^{(2)}+2\operatorname{Re}\!\left(\widetilde{\alpha}_{\ell}\vec{y}_{\ell}^{(2)}\right)$
\EndFor
\State $\bm{n}^{(0)} \gets \alpha_0\bm{n}^{(0)}$
\State $\vec{u}_{i}^{(\mathrm{n},x)} \gets \alpha_0\vec{u}_{i}^{(\mathrm{n},x)}$ for each active $(i,x)$
\State $\delta\bm{n}^{(2)} \gets \alpha_0\delta\bm{n}^{(2)}$
\State \Return $\bm{n}^{(0)}+\delta\bm{n}^{(2)}$
\end{algorithmic}
\end{algorithm}

\begin{algorithm}
\caption{Description of unmodified $2L$-th order CRAM for treatment of the Bateman equations.}
\label{alg:oldcram}
\begin{algorithmic}[1]
\State $\bm{n}^{(0)} \gets \bm{n}_0$
\For{$\ell \gets 1$ to $L$}
    \State Solve $\left(tQ(\hat{r})-\theta_{\ell}\mat{I}_K\right)\vec{y}_{\ell}=\bm{n}^{(0)}$ for $\vec{y}_{\ell}$
    \State $\bm{n}^{(0)} \gets \bm{n}^{(0)}+2\operatorname{Re}\!\left(\widetilde{\alpha}_{\ell}\vec{y}_{\ell}\right)$
\EndFor
\State $\bm{n}^{(0)} \gets \alpha_0\bm{n}^{(0)}$
\State \Return $\bm{n}^{(0)}$
\end{algorithmic}
\end{algorithm}

\section{Demonstration Applications}
\label{sec:demonstration}

\subsection{Toy Problem Application}
\label{sec:toy}

A controlled ten-nuclide toy problem was constructed to reproduce qualitative features of a depletion matrix while keeping the reaction network small enough to interpret directly.
This problem serves as a demonstration case and is not a physical depletion chain.
A complete description of this problem is provided to allow readers to verify their own implementations.
It combines a sparse transmutation structure, correlated stochastic reaction rates, fission-like multi-product channels, and a small cyclic component in the matrix.

\begin{figure}[h]
  \centering
  \safeincludegraphics[width=0.92\columnwidth]{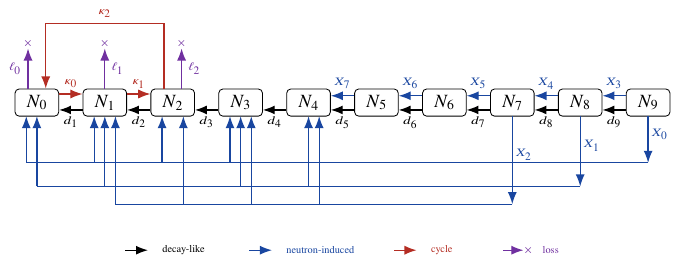}
  \caption{Schematic of the ten-nuclide toy problem.}
  \label{fig:toy_scheme}
\end{figure}

An initial composition of
\begin{equation}
  \vec{n}_0
  =
  \begin{bmatrix}
  0 & 0 & 0 & 0 & 0 & 0.10 & 0.15 & 0.20 & 0.25 & 0.30
  \end{bmatrix}^{T}
  \mathrm{atoms\,cm^{-3}}
  \label{eq:toy_n0}
\end{equation}
is evolved over a period of $t=10\,\mathrm{s}$ according to the reaction network described in Figure~\ref{fig:toy_scheme}.
Some of the transmutations present in this toy problem are subject to uncertainty that would theoretically arise from the use of a Monte Carlo neutron transport code.
The transmutation rates that are not subject to uncertainty are then summarized in Table~\ref{tab:toy_deterministic_channels}, and those with uncertainty are listed separately in Table~\ref{tab:toy_channels}.
More specifically, a set of 9 uncertainty-free nearest-neighbor transition rates, can be specified with
\begin{equation}
  d_j = \left(0.004 + 0.0003\,j\right)\,\mathrm{s}^{-1},
  \qquad
  j=1,\ldots,9
  \label{eq:toy_decaylike}
\end{equation}
where $d_j$ represents the transmutation rate of $N_j$ to $N_{j-1}$.
Then,  a closed cycle between $N_0$, $N_1$ and $N_2$ is driven by transmutation rates $\kappa_0$, $\kappa_1$ and $\kappa_2$ all equal to $0.01\,\mathrm{s}^{-1}$.
This cycle is introduced to create a small complex-conjugate eigenvalue pair to mimic those seen in full transmutation matrices.
Additional fixed losses were assigned to $N_0$, $N_1$, and $N_2$ with rates $\ell_0 = 0.0010\,\mathrm{s}^{-1}$, $\ell_1 = 0.0008\,\mathrm{s}^{-1}$, and $\ell_2 = 0.0006\,\mathrm{s}^{-1}$, respectively.
These losses do not contribute to other isotopic concentrations.
Within the formalism of Section~\ref{sec:matrex_bateman}, the rates $d_j$, $\kappa_j$, and $\ell_j$ are treated as decay-rates and are incorporated into $\mat{\Lambda}$.
Specifically, $d_j=\lambda_{j\rightarrow j-1}$, $\kappa_0=\lambda_{0\rightarrow1}$, $\kappa_1=\lambda_{1\rightarrow2}$, and $\kappa_2=\lambda_{2\rightarrow0}$, while each $\ell_j$ contributes only $-\ell_j$ to $\left(\mat{\Lambda}\right)_{j,j}$.

\begin{table}[!htbp]
  \centering
  \caption{Deterministic transition channels used in the ten-nuclide toy
           problem.}
  \label{tab:toy_deterministic_channels}
  \small
  \begin{tabular}{lll}
    \toprule
    Channel & Transition & Rate ($\mathrm{s}^{-1}$) \\
    \midrule
    \multicolumn{3}{l}{\textit{Decay-like transitions}} \\
    $d_j$,\; $j=1,\ldots,9$
      & $N_j \rightarrow N_{j-1}$
      & $0.004 + 0.0003\,j$ \\
    \addlinespace
    \multicolumn{3}{l}{\textit{Cycle transitions}} \\
    $\kappa_0$
      &{$N_0 \rightarrow N_1$}
      & $0.0100$ \\
    $\kappa_1$
      & {$N_1 \rightarrow N_2$}
      & $0.0100$ \\
    $\kappa_2$
      & {$N_2 \rightarrow N_0$}
      & $0.0100$ \\
    \addlinespace
    \multicolumn{3}{l}{\textit{External sink terms}} \\
    $\ell_0$
      &{$N_0 \rightarrow \text{external}$}
      & $0.0010$ \\
       $\ell_1$
      & {$N_1 \rightarrow \text{external}$}
      & $0.0008$ \\
       $\ell_2$
      &{$N_2 \rightarrow \text{external}$}
      & $0.0006$ \\
    \addlinespace
    \bottomrule
  \end{tabular}
\end{table}

\begin{table*}[!htbp]
  \centering
\caption{Stochastic neutron-induced channels in the ten-nuclide toy problem. Here, $\alpha$ is used to identify each reaction channel rather than the $i$ and $x$ notation used in Section~\ref{sec:rrq}.}
  \label{tab:toy_channels}
  \small
  \begin{tabular}{cc>{\raggedright\arraybackslash}p{0.50\textwidth}cc}
    \toprule
    Channel & Parent & Product Yields & $r_\alpha$ ($\mathrm{s}^{-1}$)  & $s_\alpha$ ($\mathrm{s}^{-1}$)  \\
    \midrule
    $X_0$ & $N_9$ & $0.75N_0+0.65N_1+0.35N_2+0.15N_3$ & 0.90 & 0.09 \\
    $X_1$ & $N_8$ & $0.55N_0+0.50N_1+0.30N_3+0.10N_4$ & 0.76 & 0.08 \\
    $X_2$ & $N_7$ & $0.45N_1+0.40N_2+0.25N_3+0.12N_4$ & 0.60 & 0.06 \\
    $X_3$ & $N_9$ & $N_8$                               & 0.36 & 0.04 \\
    $X_4$ & $N_8$ & $N_7$                               & 0.30 & 0.03 \\
    $X_5$ & $N_7$ & $N_6$                               & 0.24 & 0.02 \\
    $X_6$ & $N_6$ & $N_5$                               & 0.20 & 0.02 \\
    $X_7$ & $N_5$ & $N_4$                               & 0.16 & 0.02 \\
    \bottomrule
  \end{tabular}
\end{table*}

Then, there are eight reaction channels whose reaction rates are subject to uncertainty.
The first three channels, with rates $X_0$, $X_1$, and $X_2$, are fission-like multi-product reactions, in which one parent nuclide produces several lower-index nuclides.
The remaining channels, with rates $X_3$ through $X_7$, are single-product transmutation reactions.
The reaction rates corresponding to these channels follow a multivariate normal distribution, with exact/mean reaction rates $r_i^{(\mathrm{n},x)}$, standard deviations $s_i^{(\mathrm{n},x)}$, and correlations $\rho_{ix,jy}$ as defined in Section~\ref{sec:rrq}.
For this simplified toy problem, each combination of the indices $i$ and $x$ corresponds to a single reaction channel.
The exact reaction rates and standard deviations are provided in Table~\ref{tab:toy_channels}, while the construction and numerical values of the correlations are provided in \ref{app:toy_correlation_matrix}.
These distribution parameters are directly used in the second-order correction calculation and in the stochastic sampling described in Section~\ref{sec:toy_results}.

\subsubsection{Representative Features of Toy Problem}
\label{sec:toy_sparsity}

\begin{figure*}[h]
  \centering
  \safeincludegraphics[width=0.96\textwidth]{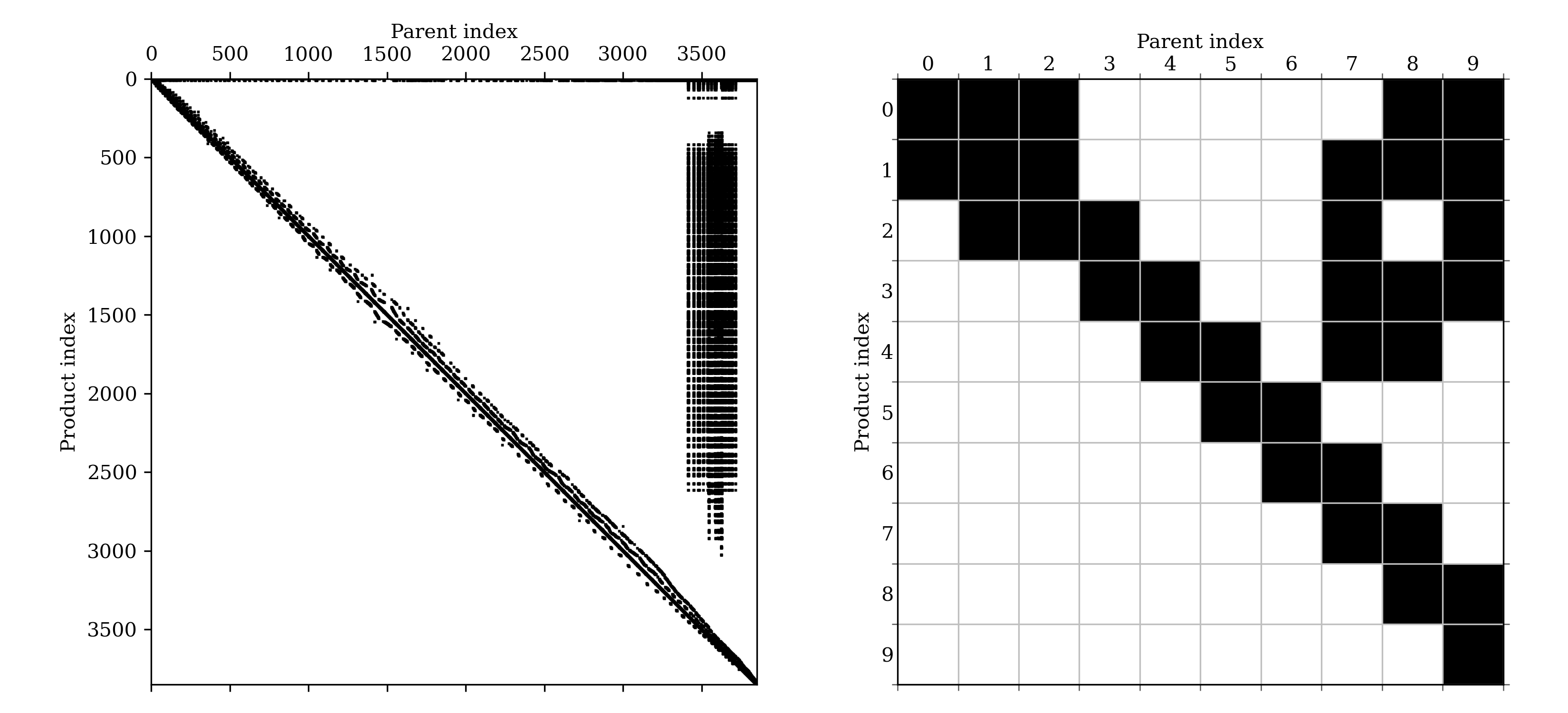}
  \caption{Sparsity patterns of representative UO$_2$ depletion matrix and the
  ten-nuclide toy matrix. A filled cell represents a nonzero coupling between a parent nuclide $N_j$ (column index) and a product or daughter nuclide $N_i$ (row index).}
  \label{fig:sparsity}
\end{figure*}

As noted above, the toy matrix is not intended to reproduce an actual physical system.
Instead, it preserves the qualitative behavior of a typical transmutation matrix.
Figure~\ref{fig:sparsity} compares the sparsity pattern of a representative  depletion matrix extracted for a UO$_2$ 3.5\% fuel with that of the ten-nuclide toy problem. 
The representative UO$_2$ matrix has dimension $3851\times3851$ with 83,644 nonzero entries, whereas the toy matrix has dimension $10\times10$ with $34$ nonzero entries.
Generally, the near-diagonal entries tend to describe decay and neutron absorption reactions while entries corresponding to fission reactions occupy high-column, lower-row indices.

\begin{figure*}[!htbp]
  \centering
  \safeincludegraphics[width=0.96\textwidth]{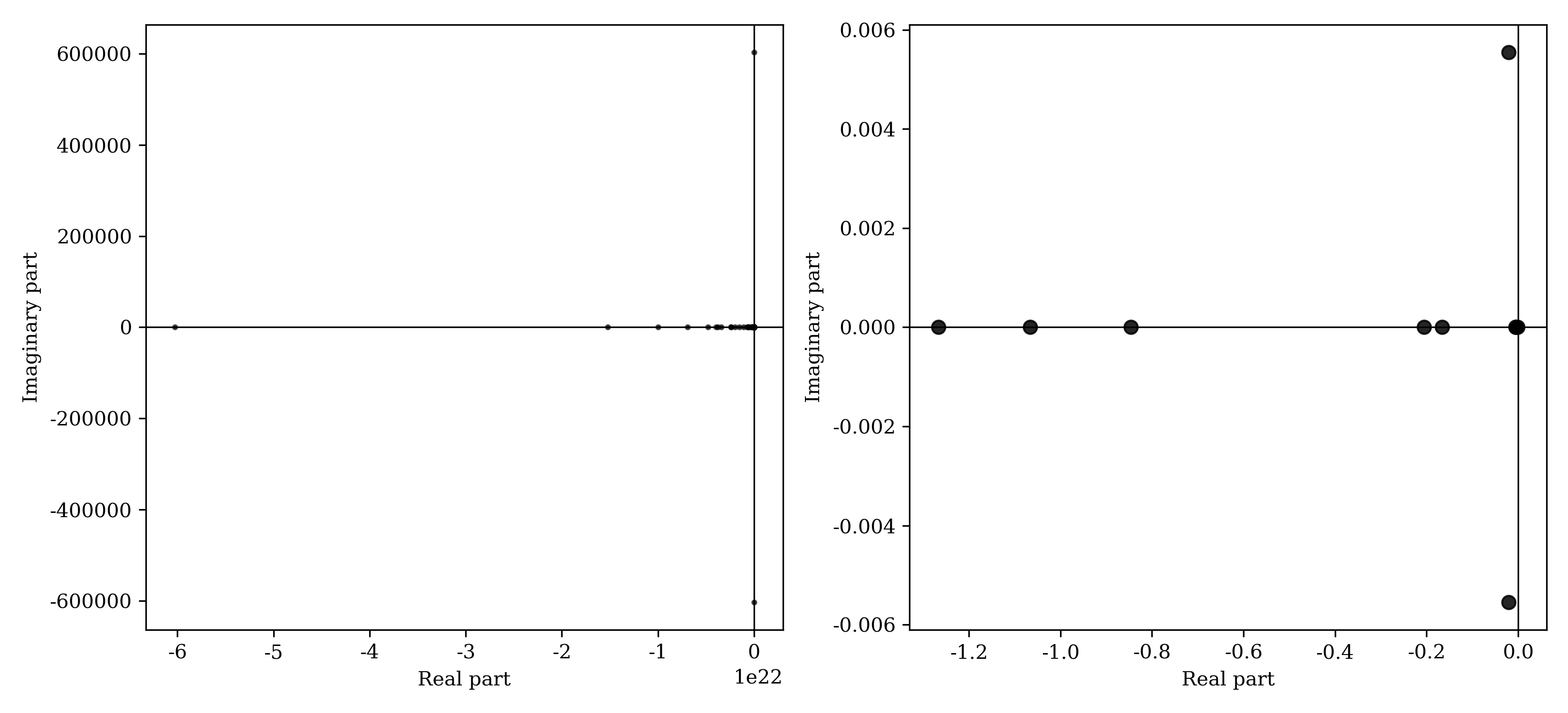}
  \caption{Eigenvalue distributions of the representative UO$_2$ depletion matrix and the
  ten-nuclide toy matrix.}
  \label{fig:eigs}
\end{figure*}

Figure~\ref{fig:eigs} shows the eigenvalue distribution for the representative UO$_2$ and toy problem transmutation matrices. For the representative UO$_2$ matrix, the 100 dominant eigenvalues (largest magnitude) were extracted.
Such eigenvalue plots are useful for comparing the spectral structure of burnup matrices, since the eigenvalue distribution is closely related to the stiffness of the Bateman system and to the numerical behavior of matrix-exponential solvers such as CRAM. 
Resembling those of the UO$_2$ matrix, the toy matrix contains mostly negative eigenvalues with one complex-conjugate pair generated by the imposed cycle.

\FloatBarrier
\subsubsection{Method Comparison Against an Exact-Rate Reference}
\label{sec:toy_results}

The second-order correction was assessed against an exact-rate reference. Since the exact reaction rates $\vec{r}$ are known for this controlled problem, the reference concentration vector, $\vec{n}_{\mathrm{ref}}$, can be directly obtained.
To investigate the bias of calculated isotope vectors, a sampling approach is used where a total of $M=20{,}000$ reaction-rate realizations were independently sampled.
These realizations were generated as $10{,}000$ independent antithetic pairs, using opposite perturbations about $\vec{r}$, to reduce the statistical variance of the estimated mean without changing its expectation.
For each realization $m$, the conventional Bateman solution is evaluated using the sampled reaction rates to form $\vec{n}^{(0,m)}$.
This can then be bias-corrected with $\vec{n}^{(2,m)} = \vec{n}^{(0,m)}+\delta\vec{n}^{(2,m)}$.
The mean concentration vectors across all samples can be represented with $\overline{\bm{n}}^{(0)}$ and $\overline{\bm{n}}^{(2)}$.

Table~\ref{tab:toy_results} compares the uncorrected and second-order corrected ensemble means with the exact-rate reference at $t=10\,\mathrm{s}$.
The correction nearly eliminates the estimated bias for $N_0$ through $N_6$ and substantially reduces it for $N_7$ through $N_9$.
For these latter nuclides, the bias changes from $22.82\%$ to $-2.80\%$ for $N_7$, from $47.60\%$ to $-10.25\%$ for $N_8$ and from $86.79\%$ to $-29.86\%$ for $N_9$.
The remaining discrepancy is attributed to higher-order contributions that are not included in the second-order correction.
It can also be noted that the large relative bias values in $N_7$, $N_8$ and $N_9$ are amplified by their small concentrations after depletion.
These results should be taken as a demonstration of the methodology and not as an indication of  performance gains in real applications.

\begin{table*}[!htbp]
  \centering
  \caption{Comparison between the ensemble-averaged uncorrected and second-order corrected estimates and the exact-rate reference at $t=10\,\mathrm{s}$.}
  \label{tab:toy_results}
  \small
  \begin{tabular}{lccccc}
    \toprule
    Nuclide & $\overline{\vec{n}}^{(0)}$ & $\overline{\vec{n}}^{(2)}$
            & $\vec{n}_{\mathrm{ref}}$
            & Bias without correction (\%) & Bias with correction (\%) \\
    \midrule
    $N_0$ & $2.9065{\times}10^{-1}$ & $2.9085{\times}10^{-1}$
          & $2.9084{\times}10^{-1}$ & $-0.0652$ & $0.0026$ \\
    $N_1$ & $3.3965{\times}10^{-1}$ & $3.3985{\times}10^{-1}$
          & $3.3984{\times}10^{-1}$ & $-0.0545$ & $0.0028$ \\
    $N_2$ & $1.7382{\times}10^{-1}$ & $1.7393{\times}10^{-1}$
          & $1.7392{\times}10^{-1}$ & $-0.0557$ & $0.0029$ \\
    $N_3$ & $1.5715{\times}10^{-1}$ & $1.5727{\times}10^{-1}$
          & $1.5726{\times}10^{-1}$ & $-0.0694$ & $0.0034$ \\
    $N_4$ & $2.4333{\times}10^{-1}$ & $2.4448{\times}10^{-1}$
          & $2.4447{\times}10^{-1}$ & $-0.4659$ & $0.0047$ \\
    $N_5$ & $9.9259{\times}10^{-2}$ & $9.8546{\times}10^{-2}$
          & $9.8547{\times}10^{-2}$ & $0.7226$ & $-0.0009$ \\
    $N_6$ & $3.5849{\times}10^{-2}$ & $3.5065{\times}10^{-2}$
          & $3.5088{\times}10^{-2}$ & $2.1690$ & $-0.0656$ \\
    $N_7$ & $2.0649{\times}10^{-4}$ & $1.6340{\times}10^{-4}$
          & $1.6812{\times}10^{-4}$ & $22.8225$ & $-2.8047$ \\
    $N_8$ & $2.5014{\times}10^{-5}$ & $1.5210{\times}10^{-5}$
          & $1.6947{\times}10^{-5}$ & $47.5984$ & $-10.2493$ \\
    $N_9$ & $1.7671{\times}10^{-6}$ & $6.6355{\times}10^{-7}$
          & $9.4605{\times}10^{-7}$ & $86.7880$ & $-29.8611$ \\
    \bottomrule
  \end{tabular}
\end{table*}

\subsubsection{Correction Growth with Reaction-Rate Uncertainty}
\label{sec:toy_bias_growth}

\begin{figure*}[!htbp]
  \centering
  \safeincludegraphics[width=0.90\textwidth]{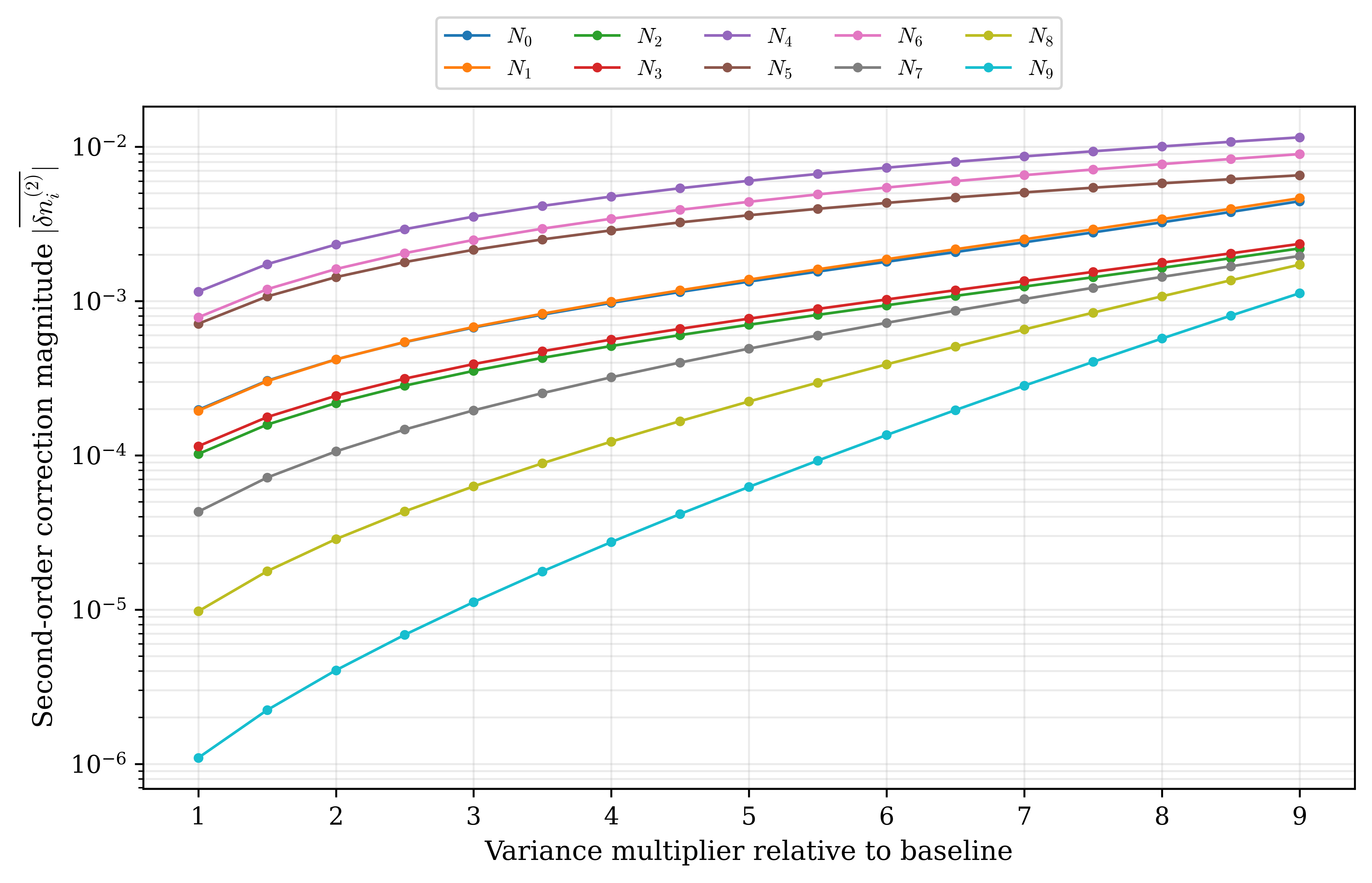}
  \caption{Magnitude of the second-order correction as a function of the reaction-rate variance multiplier in the ten-nuclide toy problem.}
  \label{fig:bias_growth}
\end{figure*}

Figure~\ref{fig:bias_growth} shows how the second-order correction magnitude varies with reaction-rate uncertainty.
The variance of all stochastic channels was scaled uniformly by a factor of $k$ while keeping the reaction rates and correlations fixed.
For each value of $k$, the correction magnitude is estimated using $M=20{,}000$ reaction-rate realizations generated as $10{,}000$ independent antithetic pairs, consistent with the sampling procedure of Section \ref{sec:toy_results}.

The magnitude of the correction increases with the variance multiplier for all nuclides, showing that larger reaction-rate uncertainty produces a larger second-order contribution to the depletion solution.
The growth is nonlinear because increasing $k$ broadens the sampled reaction-rate distribution; the correction is evaluated at each sampled rate vector and therefore involves depletion matrices and sensitivities that also vary with $k$.
Nuclides $N_7$, $N_8$ and $N_9$ exhibit the most pronounced correction growth as the uncertainty increases. 
This stronger dependence on $k$ is consistent with their concentrations being directly controlled by the largest stochastic removal rates in the system.
More generally, the nonlinear dependence on $k$ arises because increasing the variance both scales the covariance entering the correction and broadens the sampled reaction-rate distribution, causing the depletion matrices and associated second-order sensitivities to vary with the uncertainty level.

\FloatBarrier
\subsection{Deep Penetration Application}
\label{sec:tp2}

The second demonstration considers a deep-penetration activation problem based on the ITER port-plug model of ~\citet{romano2025implementation}.
Unlike the controlled toy problem of Section \ref{sec:toy}, the stochastic reaction rates are obtained directly from Monte Carlo particle transport.
A single constant-matrix depletion step is considered here to isolate the behavior of the bias correction.

\begin{figure*}[t]
    \centering
    \safeincludegraphics[width=0.78\textwidth]{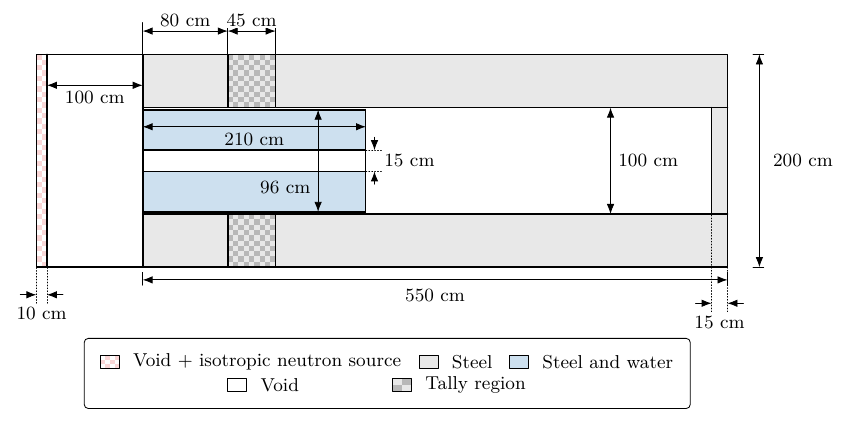}
    \caption{Schematic of the ITER port-plug model and selected reaction-rate tally region, based on the configuration of \citet{romano2025implementation}.}
    \label{fig:tp2_geometry}
\end{figure*}

As shown in Figure \ref{fig:tp2_geometry}, the model consists of a $550$-cm-long cylindrical port plug with inner and outer radii of $50$ and $100$~cm.
The inner section contains a $210$-cm insert composed of $80$~vol.\% 316 stainless steel and $20$~vol.\% water.
An isotropic monoenergetic $14$-MeV neutron source is located outside of this port plug.
Reaction rates are tallied in the annular region $50<r<100$~cm and $80<z<125$~cm, where neutron attenuation produces significant Monte Carlo uncertainty.
A reduced transmutation chain with $K=520$ nuclides is used for the reaction families $(n,x)$, with $x\in\{\gamma,\,2n,\,3n,\,4n,\,p,\,\alpha\}$, and 779 statistically active nuclide--reaction channels are retained.

The irradiation history of \citet{romano2025implementation} is represented by a time-averaged source strength of approximately $7.6\times10^{17}$~n\,s$^{-1}$ over $t=14$ years.
For this numerical demonstration, the source strength is multiplied by $10^7$ to amplify the second-order effect.
It does not represent an ITER operating condition.

\subsubsection{Reaction-Rate Statistics}
\label{sec:tp2_rate_statistics}

\begin{figure*}[t]
    \centering
    \safeincludegraphics[width=0.78\textwidth]{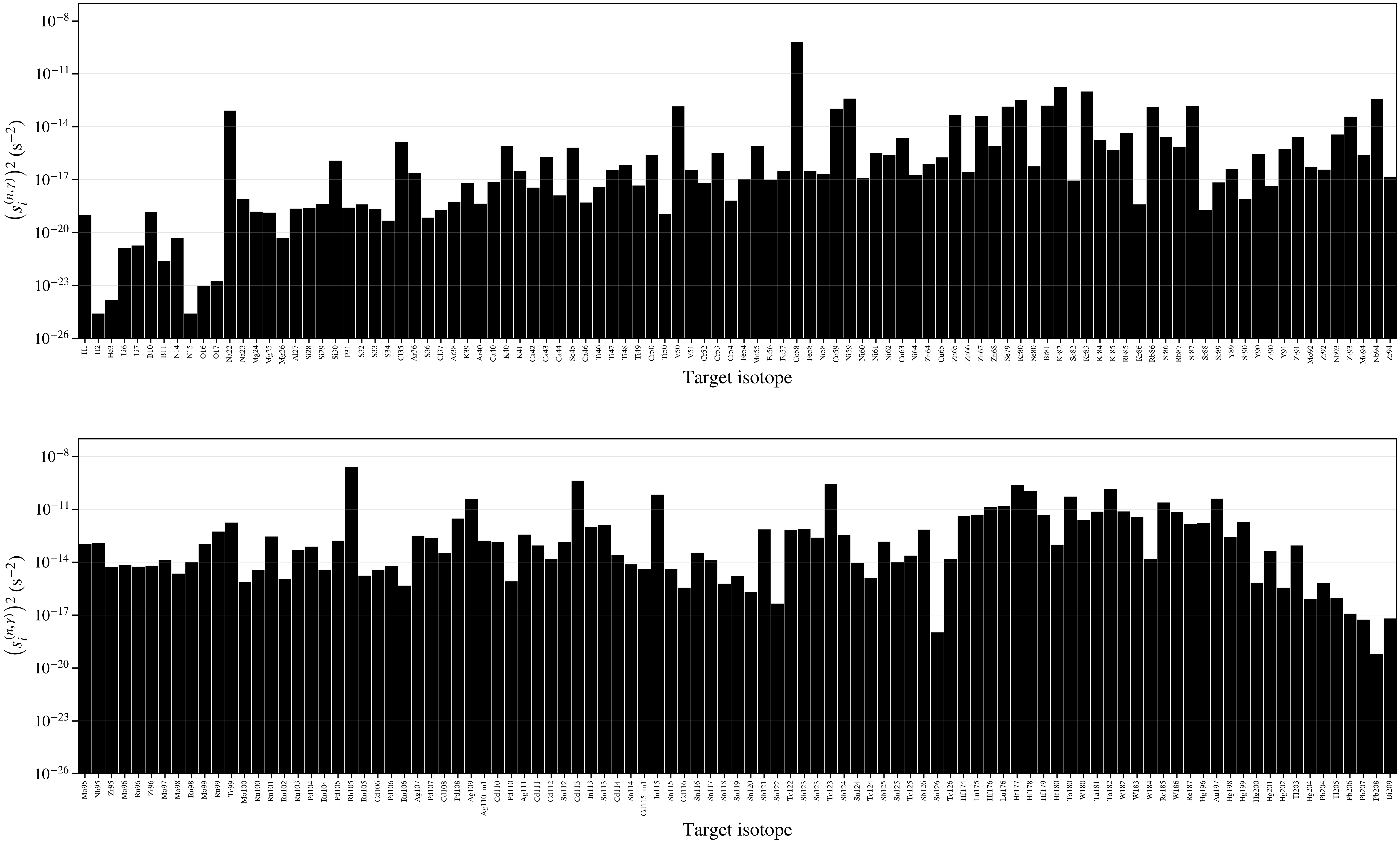}
    \caption{Variance $\left(\hat{s}_{i}^{(\mathrm{n},\gamma)}\right)^2$ of the nuclide-specific $(n,\gamma)$ reaction-rate estimators in the selected port-plug region.}
    \label{fig:tp2_ngamma_variance}
\end{figure*}

\begin{figure*}[t]
    \centering
    \safeincludegraphics[width=0.8\textwidth]{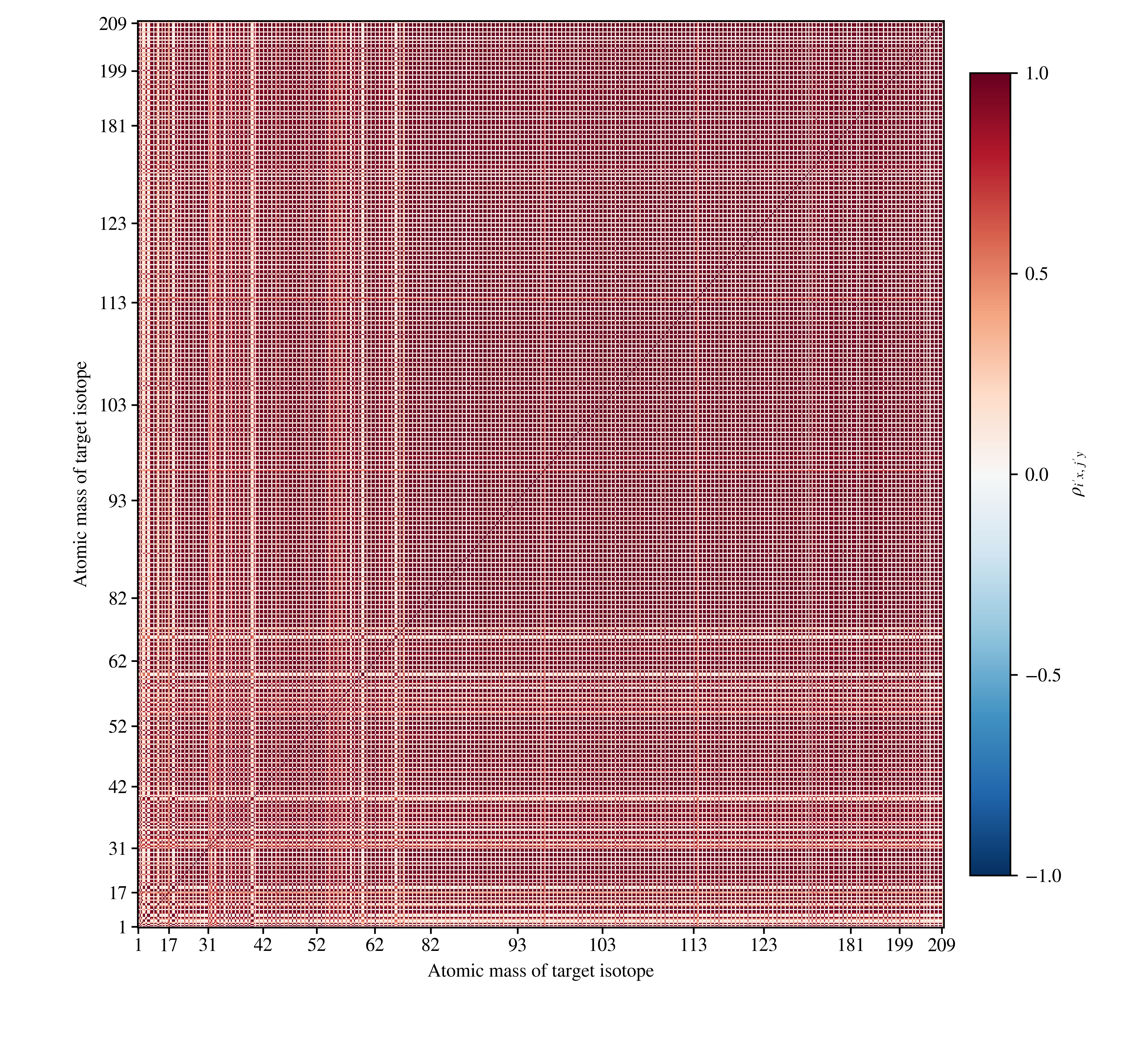}
    \caption{Estimated correlation coefficients $\hat{\rho}_{ix,jy}$ between the 779 active nuclide--reaction channels.}
    \label{fig:tp2_correlation}
\end{figure*}

Evaluation of $\bm{n}^{(2)}(t;\hat{r})$ requires the reaction-rate uncertainty estimates $\hat{s}_{i}^{(\mathrm{n},x)}$ and $\hat{\rho}_{ix,jy}$ in addition to the reaction-rate estimates $\hat{r}$.
Unlike the toy problem, these quantities must be estimated from the Monte Carlo transport data.
Table~\ref{tab:tp2_rates} summarizes the reaction-rate statistics by reaction type for the limited-statistics (LS) calculation used below.
This LS calculation relied on $100$ batches with $10^5$ particles within each batch.
The uncertainty varies substantially between individual nuclide--reaction channels and the $(\mathrm{n}, 4\mathrm{n})$ channel was not activated for any history.
Figure~\ref{fig:tp2_ngamma_variance} illustrates this behavior for the dominant $(n,\gamma)$ type, with the remaining reaction families shown in~\ref{app:tp2_additional_variances}.
The reaction-rate estimators also exhibit a complex covariance structure.
Figure~\ref{fig:tp2_correlation} shows the estimated correlation coefficients, $\hat{\rho}_{ix,jy}$, between the 779 active channels.
The pronounced off-diagonal structure demonstrates the importance of retaining correlations in the second-order correction.
These correlations arise from the implicit sampling approach that allows the same histories to contribute to multiple reaction channels.

\begin{table*}[t]
    \centering
    \caption{Reaction-rate statistics by reaction type for the LS calculation.}
    \label{tab:tp2_rates}
    \small
    \begin{tabular}{lrrrr}
        \toprule
        Reaction type
        & Active channels
        & Mean rate ($\mathrm{s}^{-1}$)
        & Std.\ dev.\ ($\mathrm{s}^{-1}$)
        & Relative std.\ (\%) \\
        \midrule
        $(\mathrm{n},\gamma)$
        & 199
        & $3.68\times10^{-4}$
        & $1.60\times10^{-5}$
        & $4.35$ \\
        $(\mathrm{n},2\mathrm{n})$
        & 190
        & $2.46\times10^{-7}$
        & $2.33\times10^{-8}$
        & $9.46$ \\
        $(\mathrm{n},3\mathrm{n})$
        & 7
        & $8.12\times10^{-11}$
        & $7.87\times10^{-12}$
        & $9.68$ \\
        $(\mathrm{n},4\mathrm{n})$
        & 0
        & $0$
        & $0$
        & -- \\
        $(\mathrm{n},\mathrm{p})$
        & 192
        & $1.05\times10^{-4}$
        & $3.87\times10^{-6}$
        & $3.68$ \\
        $(n,\alpha)$
        & 191
        & $1.04\times10^{-5}$
        & $3.79\times10^{-7}$
        & $3.64$ \\
        \bottomrule
    \end{tabular}
\end{table*}

\FloatBarrier

\subsubsection{Bias Correction Results}
\label{sec:tp2_results}

A high-statistics (HS) calculation with $10^6$ batches and $10^5$ particles within each batch is generated and used to construct the numerical reference.
To investigate the impact of errors in $\hat{s}_i^{(\mathrm{n}, x)}$ and $\hat{\rho}_{ix,jy}$ on the correction, $\bm{n}^{(2)}$  is evaluated using two approaches
\begin{enumerate}
\item HS $\bm{n}^{(2)}$: Although $\hat{r}$ are obtained from each LS calculation, both $\hat{\rho}_{ix,jy}$ and $\hat{s}_i^{(\mathrm{n}, x)}$ are estimated using the HS calculation and then used to correct $\bm{n}^{(0)}$. In practice, this is not a viable approach as the estimation of  $\bm{n}^{(2)}$ would be using information from a calculation with better statistics. 
\item $\bm{n}^{(2)}$: $\hat{\rho}_{ix,jy}$ and $\hat{s}_i^{(\mathrm{n}, x)}$ are estimated using information from only the LS calculation.
This represents the realistic scenario where an analyst is computing $\bm{n}^{(2)}$ based on the same LS calculation.
\end{enumerate}
Table~\ref{tab:tp2_nuclides_b100} compares the resulting concentration biases for five representative nuclides.
These results are generated by partitioning the $10^{6}$ batches into $10^{4}$ nonoverlapping LS calculations of $100$ batches, with $10^{5}$ particles within each batch.
This paired construction preserves the sampling distribution of each LS calculation while causing the first-order contribution of reaction-rate fluctuations to cancel in the comparison.
The resulting uncorrected and corrected isotope vectors are averaged over all LS calculations and compared with the isotope vector obtained from the HS calculation.

Both covariance treatments substantially reduce the bias, with the locally estimated covariance reducing the bias by factors between approximately $42$ and $69$.
The corresponding global inventory bias is summarized in Table~\ref{tab:tp2_total_b100}.
The second-order correction reduces the normalized global $L_1$ bias by factors of $121.3$ and $51.4$ using the HS and local covariance estimates, respectively.

\begin{table*}[t]
    \centering
    \caption{Absolute relative concentration biases for selected nuclides before and after the second-order correction. Reference  concentration was obtained from HS calculation.}
    \label{tab:tp2_nuclides_b100}
    \small
    \begin{tabular}{lrrr}
        \toprule
        Nuclide
        & $\overline{\vec{n}}^{(0)}$ bias (\%)
        & HS $\, \, \overline{\vec{n}}^{(2)}$ bias (\%)
        & $\overline{\vec{n}}^{(2)}$ bias  (\%) \\
        \midrule
        $^{52}\mathrm{Cr}$
        & $0.58615$
        & $6.5719\times10^{-4}$
        & $0.01059$\\
        $^{58}\mathrm{Fe}$
        & $0.33077$
        & $0.003323$
        & $0.007226$ \\
        $^{56}\mathrm{Fe}$
        & $0.24908$
        & $0.001976$
        & $0.003635$ \\
        $^{57}\mathrm{Fe}$
        & $0.24913$
        & $0.004036$
        & $0.005933$\\
        $^{60}\mathrm{Co}$
        & $0.07864$
        & $3.1854\times10^{-4}$
        & $0.001212$ \\
        \bottomrule
    \end{tabular}
\end{table*}

\begin{table}[t]
    \centering
    \caption{Global inventory bias summary and bias reduction factor (BRF). Reference concentration was obtained from HS calculation.}
    \label{tab:tp2_total_b100}
    \small
    \begin{tabular}{lrrr}
        \toprule
        Metric
        & $\overline{\vec{n}}^{(0)}$
        & HS $\, \, \overline{\vec{n}}^{(2)}$
        & $\overline{\vec{n}}^{(2)}$ \\
        \midrule
        Normalized $L_1$ bias
        & $9.7847\times10^{-5}$
        & $8.0689\times10^{-7}$
        & $1.9053\times10^{-6}$ \\
        BRF
        & --
        & $121.3$
        & $51.4$ \\
        \bottomrule
    \end{tabular}
\end{table}

Overall, the deep-penetration application demonstrates that the second-order correction remains effective for a representative activation problem involving hundreds of correlated stochastic reaction channels, including when the required covariance information is estimated from the same limited Monte Carlo calculation.
\FloatBarrier

\subsection{BWR 9-Pin Application}
\label{sec:BWR-lattice}
\begin{figure*}[t]
    \centering
    \safeincludegraphics[width=0.85\textwidth]{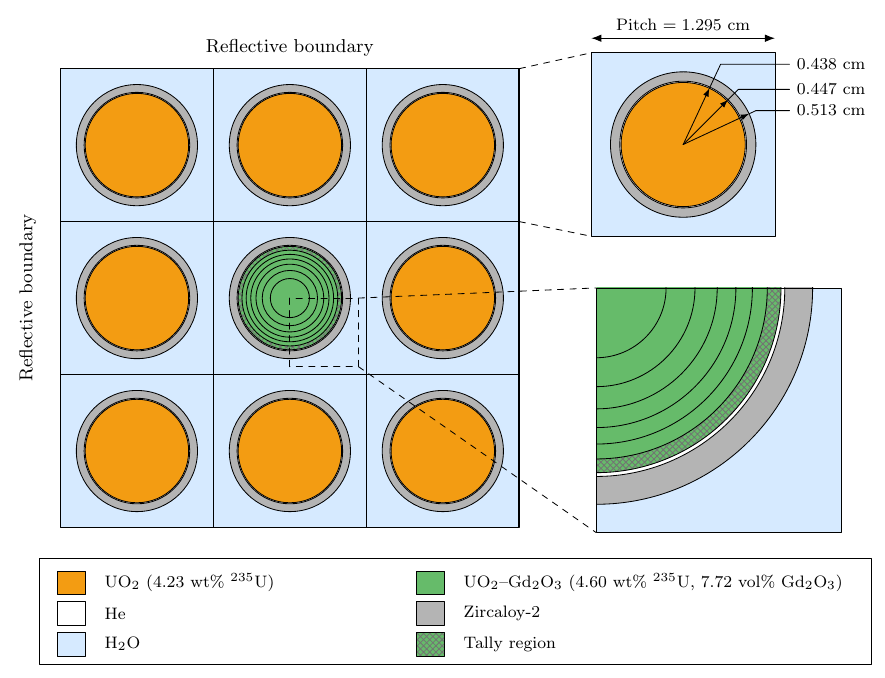}
    \caption{Schematic of a boiling water reactor (BWR) 9-pin geometry based on pin geometries and materials provided in \citet{radaideh2019advanced} as CS0.}
    \label{fig:tp3_geometry}
\end{figure*}

This final demonstration considers a BWR 9-pin geometry based on the pin geometries and materials provided in \citet{radaideh2019advanced} as CS0.
A diagram of the model is given in Figure \ref{fig:tp3_geometry}.
Again, a single constant-matrix depletion step is used with stochastic reaction rates obtained directly from Monte Carlo particle transport.
Unlike the previous demonstration application, effort is taken here to replicate a practical analysis scenario.
First, a full transmutation chain with $K=3819$ nuclides is used for the reaction rates $(n,x)$, with $x\in\{\gamma,\,2n,\,3n,\,4n,\,p,\,\alpha\}$ including fission, containing 2562 distinct nuclide-reaction channels.
Next, there is no artificial inflation of the neutron flux, a representative 25 MW/MTU power density is imposed on the complete 9-pin geometry with a time step of 2 GWd/MTU or $t = 80$ days.

Also unlike the previous demonstration application, this demonstration requires the use of an eigenvalue calculation.
Notably, in conventional eigenvalue calculation configurations, correlations are introduced between generations that would bias estimates of $\hat{s}_{i'}^{(\mathrm{n},x)}$ and $\hat{\rho}_{ix,jy}$.
As such, in this section, each sampled generation is preceded by 1500 inactive generations to obtain a fission source distribution independent of those used for the other samples.
Although computationally inefficient, this approach enables verification of the methodology in the absence of intergenerational correlations.

\FloatBarrier

\subsubsection{Bias Correction Results}
To explore the impact of the second-order bias correction, $5 \times 10^4$ batches were generated with $10^3$ particles within each batch.
These $5 \times 10^4$ batches formed the HS calculation for this demonstration application while sets of non-overlapping 25 batches formed the 2000 LS calculations.
A summary of the expected reaction rate uncertainties is provided in Table \ref{tab:reaction_rate_statistics3} based on this set of 2000 LS calculations.
Then Table \ref{tab:nuclide_errors3} shows the uncorrected and corrected bias estimations for 11 selected isotopes and finally, Table \ref{tab:inventory_errors3} shows an aggregate measure of the uncorrected and corrected isotope concentrations.
These results again demonstrate the typical reduction in bias offered by the second-order correction.
Here, Pu-242 and Cm-244 are included in this table to show that this statistical verification approach does not show a reduction in the bias of every nuclide.
Statistical uncertainty in $\hat{s}_{i'}^{(\mathrm{n},x)}$ and $\hat{\rho}_{ix,jy}$ can cause the correction to increase the bias for some nuclides.
The finite number of realizations used to generate the results in this table also introduces uncertainty to these results.
For reference, using typical Monte Carlo scaling, approximately 140 batches are estimated to be required to match the $L_1$ bias of a corrected 25-batch calculation.
This also leads to a 5.5 theoretical transport-cost reduction factor for the same bias level.

\begin{table*}[t]
    \centering
    \caption{Reaction-rate statistics by reaction type for a 25-batch LS calculation.}
    \label{tab:reaction_rate_statistics3}
    \small
    \begin{tabular}{lrrrr}
        \toprule
        Reaction type & Active channels & Mean rate ($\mathrm{s}^{-1}$) & Std.\ dev.\ ($\mathrm{s}^{-1}$) & Relative std.\ (\%) \\
        \midrule
        $(\mathrm{n},\mathrm{\gamma})$ & 553 & $1.09\times10^{-5}$ & $1.91\times10^{-7}$ & $1.75$ \\
        $(\mathrm{n},2\mathrm{n})$ & 548 & $3.92\times10^{-10}$ & $4.00\times10^{-11}$ & $10.21$ \\
        $(\mathrm{n},3\mathrm{n})$ & 393 & $6.02\times10^{-13}$ & $7.78\times10^{-13}$ & $129.16$ \\
        $(\mathrm{n},4\mathrm{n})$ & 58 & $7.54\times10^{-16}$ & $3.67\times10^{-15}$ & $486.85$ \\
        $(\mathrm{n},\mathrm{p})$ & 462 & $8.00\times10^{-7}$ & $1.55\times10^{-8}$ & $1.93$ \\
        $(\mathrm{n},\mathrm{\alpha})$ & 460 & $5.77\times10^{-8}$ & $8.07\times10^{-10}$ & $1.40$ \\
        Fission & 88 & $6.31\times10^{-7}$ & $9.18\times10^{-9}$ & $1.46$ \\
        \bottomrule
    \end{tabular}
\end{table*}

 \begin{table*}[t]
     \centering
     \caption{Absolute estimated relative concentration biases for 11 selected nuclides before and after the second-order correction. Reference concentrations were obtained from the HS calculation.}
     \label{tab:nuclide_errors3}
     \small
     \begin{tabular}{lrr}
         \toprule
         Nuclide & $\overline{\vec{n}}^{(0)}$ abs.\ bias (\%) & $\overline{\vec{n}}^{(2)}$ abs.\ bias (\%) \\
         \midrule
         $^{155}\mathrm{Gd}$ & $0.002250$ & $1.7252\times10^{-5}$  \\
         $^{156}\mathrm{Gd}$ & $0.001026$ & $7.8753\times10^{-6}$  \\
         $^{157}\mathrm{Gd}$ & $0.03577$ & $5.6752\times10^{-4}$  \\
         $^{158}\mathrm{Gd}$ & $0.006376$ & $9.5850\times10^{-5}$  \\
         $^{239}\mathrm{Pu}$ & $3.4790\times10^{-4}$ & $2.2888\times10^{-4}$  \\
         $^{241}\mathrm{Pu}$ & $0.05832$ & $2.9493\times10^{-5}$  \\
         $^{155}\mathrm{Eu}$ & $0.01960$ & $0.001301$  \\
         $^{241}\mathrm{Am}$ & $0.04530$ & $3.9710\times10^{-4}$  \\
         $^{242}\mathrm{Cm}$ & $0.01682$ & $0.007046$  \\
         $^{244}\mathrm{Cm}$ & $0.01168$ & $0.05831$  \\
         $^{242}\mathrm{Pu}$ & $0.001415$ & $0.008948$  \\
         \bottomrule
     \end{tabular}
\end{table*}

\begin{table}[t]
    \centering
    \caption{Global inventory bias summary and BRF. Reference concentrations were obtained from the HS calculation.}
    \label{tab:inventory_errors3}
    \small
    \begin{tabular}{lrr}
        \toprule
        Metric & $\overline{\vec{n}}^{(0)}$ & $\overline{\vec{n}}^{(2)}$ \\
        \midrule
        Normalized $L_1$ bias &$1.8624\times10^{-6}$& $3.2948\times10^{-8}$ \\
        BRF & -- & $56.5$ \\
        \bottomrule
    \end{tabular}
\end{table}

\FloatBarrier

\subsubsection{Computational Cost Comparison}
To evaluate the additional computational cost associated with the calculation of $\overline{\vec{n}}^{(2)}$ compared to $\overline{\vec{n}}^{(0)}$, a set of 15 LS calculations were executed on an Intel Xeon E-2388G CPU with 8 physical cores and 16 hardware threads.
Table \ref{tab:computational_cost} summarizes the measured runtime, memory usage, and linear algebra workload for both calculations.
Here, ``$N_a$, active sensitivities'' refers to the number of observed channels also represented in the depletion chain.
Although the calculation of $\overline{\vec{n}}^{(2)}$ requires substantially more wall time, the procedure described in Algorithm \ref{alg:newcram} maintains a reasonable runtime.
The use of this adapted CRAM algorithm enables factorization reuse to maintain the dense-operation $K^3$ scaling associated with conventional CRAM approaches.

\begin{table}[htbp!]
    \centering
    \caption{Computational cost comparison between the uncorrected and corrected depletion calculations.}
    \label{tab:computational_cost}
    \small
    \begin{tabular}{lrr}
        \toprule
        Metric & $\overline{\vec{n}}^{(0)}$ & $\overline{\vec{n}}^{(2)}$ \\
        \midrule
        Wall time (s) & $1.22$ & $52.8$ \\
        Maximum memory (MiB) & $388$ & $1135$ \\
        Matrix dimension & $3819$ & $3819$ \\
        $N_a$, active sensitivities & -- & $1967$ \\
        CRAM order & $48$ & $48$ \\
        Linear system size & $3819 \times 3819$ & $3819 \times 3819$ \\
        No. Linear systems & $24$ & $1969 \cdot 24$ \\ 
        Scaling & $L\left(K^3 + K^2\right)$ & $L\left(K^3 + (N_a + 2) K^2\right)$ \\
        \bottomrule
    \end{tabular}
\end{table}

\FloatBarrier

\subsubsection{Bias Versus Error Comparison}
\label{sec:bias_versus_error}
\begin{figure*}[t]
    \centering
    \safeincludegraphics[width=0.5\textwidth]{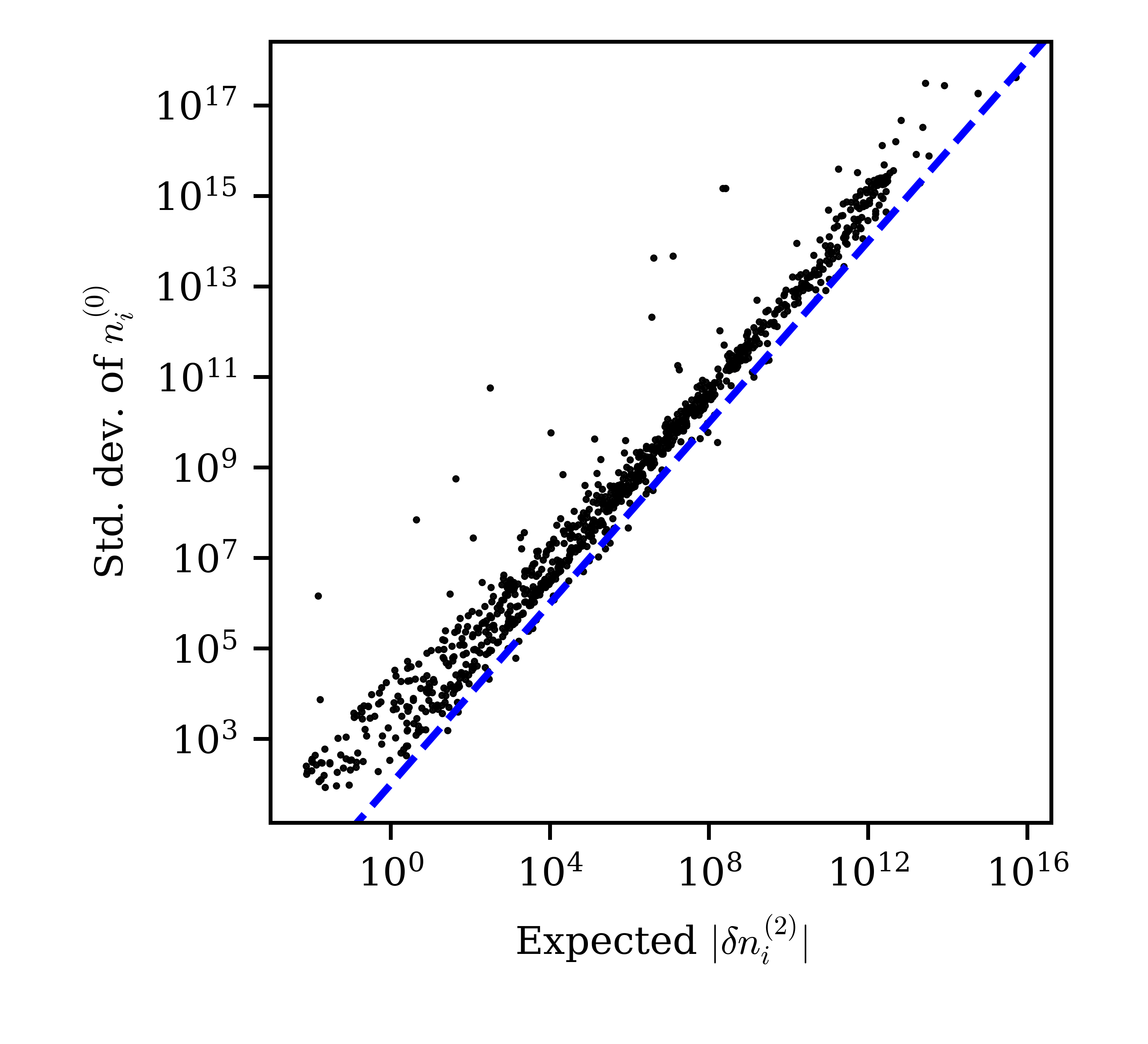}
    \caption{Across 1000 highest-concentration nuclides, standard deviation of concentration, $n_i^{(0)}$, as a function of the expected second-order correction magnitude, $\left|\delta n_i^{(2)}\right|$. The dashed line indicates where the standard deviation is 100 times the expected correction magnitude.}
    \label{fig:be_scatter}
\end{figure*}

This final results section emphasizes that the corrected estimate $\overline{\vec{n}}^{(2)}$ is intended to decrease the bias in isotope concentrations.
It is not directly intended to reduce the standard deviation of these concentrations that are induced by the stochastic reaction rates.
Figure \ref{fig:be_scatter} shows the relationship between the standard deviation of $n_i^{(0)}$ and the expected correction magnitude, $\left|\delta n_i^{(2)}\right|$, for the 1000 highest-concentration nuclides in the LS calculations.
The dashed line indicates where the standard deviation is 100 times the expected correction magnitude.
As most nuclides lie above this threshold the bias correction is generally significantly smaller than the total variability of the concentration.

Communicating this concept more directly, Table~\ref{tab:bias_versus_error} provides the average  errors of $\vec{n}^{(0)}$ and $\vec{n}^{(2)}$ across the 2000 LS calculations compared to a new reference calculation with $10^4$ batches with $10^5$ particles within each batch.
This reference calculation was generated entirely independently.
The ERFs close to unity emphasize the negligible effect of the bias correction on the total Monte Carlo-induced error in this application.
Therefore, the advantage of the bias correction lies less in reducing the error of an individual calculation than in reducing systematic offsets in the expected isotope concentrations.
For example, these systematic offsets may accumulate unpredictably across large models with multiple depletable zones or burnup steps.

\begin{table*}[t]
    \centering
    \caption{Mean absolute relative concentration errors and error reduction factors (ERF) across the LS calculations.}
    \label{tab:bias_versus_error}
    \small
    \begin{tabular}{lrrrr}
        \toprule
        Nuclide & Mean $\bm{n}^{(0)}$ error (\%) & Mean $\bm{n}^{(2)}$ error (\%) & ERF  \\
        \midrule
        $^{155}\mathrm{Gd}$ & $0.53846$ & $0.53852$ & $0.9999$ \\
        $^{156}\mathrm{Gd}$ & $0.25155$ & $0.25156$ & $1.0000$ \\
        $^{157}\mathrm{Gd}$ & $2.12278$ & $2.12341$ & $0.9997$  \\
        $^{158}\mathrm{Gd}$ & $0.37812$ & $0.37829$ & $0.9995$  \\
        $^{239}\mathrm{Pu}$ & $4.21650$ & $4.21652$ & $1.0000$  \\
        $^{241}\mathrm{Pu}$ & $7.44318$ & $7.44822$ & $0.9993$  \\
        $^{155}\mathrm{Eu}$ & $1.78378$ & $1.78483$ & $0.9994$ \\
        $^{241}\mathrm{Am}$ & $7.54066$ & $7.54508$ & $0.9994$  \\
        $^{242}\mathrm{Cm}$ & $8.28380$ & $8.28582$ & $0.9998$ \\
        $^{244}\mathrm{Cm}$ & $12.68816$ & $12.68457$ & $1.0003$ \\
        $^{242}\mathrm{Pu}$ & $8.09828$ & $8.09947$ & $0.9999$  \\
        \bottomrule
    \end{tabular}
\end{table*}

\FloatBarrier
\newpage

\section{Conclusion}
\label{sec:conc}
A second-order bias correction is derived and applied to three demonstration problems.
In addition to the mean reaction rates used to form the transmutation matrix, the correction requires estimates of their variances and correlations.
The correction is intended to mitigate systematic errors introduced when the matrix exponential is evaluated using a transmutation matrix formed from mean reaction rates.
To limit the resulting computational burden, an adapted CRAM algorithm is presented that reuses transmutation-matrix factorizations across the additional required equations to compute the correction.
One drawback of this approach is highlighted in the third demonstration problem where source iterations in the eigenvalue calculation introduce intergenerational correlations that can weaken the covariance estimates.
This study avoids these correlations at significant computational cost by preceding each sampled batch with a series of inactive source iterations.
Additionally, some discussion is provided in Section \ref{sec:bias_versus_error} to clearly distinguish the bias, addressed by this correction, and the variance in nuclide densities.
Extensions of this work to accommodate different time-integration methods \citep{josey2017development} would warrant future work but may require the estimation of cross-substep covariances.

\bibliographystyle{elsarticle-num-names} 
\bibliography{refs.bib}

\section*{Acknowledgment}
Generative artificial intelligence tools were used to assist with manuscript preparation.
This work benefited from high-performance computing resources provided by the MIT Office of Research Computing and Data (ORCD). 
Esthem Bouvet gratefully acknowledges financial support from INSTN and the scholarship program Science \& Enseignement – EDF, Institut de France et Académie des sciences.
Additionally, we would like to thank K.G. Howey for their valuable conversations surrounding this work.

\appendix

\section{Alternative Presentation of Augmented System}
\label{ap:betterpres}
The structure of $\mat{A}(\hat{r})$ as it is presented in Equation \ref{eq:adef} may not be clear.
Therefore, a less compact version can be presented as
\begin{equation}
\begingroup
\setlength{\arraycolsep}{3pt}
\resizebox{\textwidth}{!}{$
\begin{aligned}
& \mat{A}(\hat{r}) = \\
& \begin{bmatrix}
Q(\hat{r}) & \mat{0} & \mat{0} & \cdots & \mat{0} & \mat{0} & \cdots & \mat{0} & \mat{0} \\
\vec{c}_1^{(\mathrm{n},1)}\vec{e}_1^T & Q(\hat{r}) & \mat{0} & \cdots & \mat{0} & \mat{0} & \cdots & \mat{0} & \mat{0} \\
\vec{c}_2^{(\mathrm{n},1)}\vec{e}_2^T & \mat{0} & Q(\hat{r}) & \cdots & \mat{0} & \mat{0} & \cdots & \mat{0} & \mat{0} \\
\vdots & \vdots & \vdots & \ddots & \vdots & \vdots & & \vdots & \vdots \\
\vec{c}_K^{(\mathrm{n},1)}\vec{e}_K^T & \mat{0} & \mat{0} & \cdots & Q(\hat{r}) & \mat{0} & \cdots & \mat{0} & \mat{0} \\
\vec{c}_1^{(\mathrm{n},N_{\mathrm{x}})}\vec{e}_1^T & \mat{0} & \mat{0} & \cdots & \mat{0} & Q(\hat{r}) & \cdots & \mat{0} & \mat{0} \\
\vdots & \vdots & \vdots & & \vdots & \vdots & \ddots & \vdots & \vdots \\
\vec{c}_K^{(\mathrm{n},N_{\mathrm{x}})}\vec{e}_K^T & \mat{0} & \mat{0} & \cdots & \mat{0} & \mat{0} & \cdots & Q(\hat{r}) & \mat{0} \\
\mat{0} & \displaystyle -\sum_{i=1}^K\sum_x \hat{\rho}_{ix,11}\hat{s}_i^{(\mathrm{n},x)}\hat{s}_1^{(\mathrm{n},1)}\vec{c}_i^{(\mathrm{n},x)}\vec{e}_i^T & \displaystyle -\sum_{i=1}^K\sum_x \hat{\rho}_{ix,21}\hat{s}_i^{(\mathrm{n},x)}\hat{s}_2^{(\mathrm{n},1)}\vec{c}_i^{(\mathrm{n},x)}\vec{e}_i^T & \cdots & \displaystyle -\sum_{i=1}^K\sum_x \hat{\rho}_{ix,K1}\hat{s}_i^{(\mathrm{n},x)}\hat{s}_K^{(\mathrm{n},1)}\vec{c}_i^{(\mathrm{n},x)}\vec{e}_i^T & \displaystyle -\sum_{i=1}^K\sum_x \hat{\rho}_{ix,1N_{\mathrm{x}}}\hat{s}_i^{(\mathrm{n},x)}\hat{s}_1^{(\mathrm{n},N_{\mathrm{x}})}\vec{c}_i^{(\mathrm{n},x)}\vec{e}_i^T & \cdots & \displaystyle -\sum_{i=1}^K\sum_x \hat{\rho}_{ix,KN_{\mathrm{x}}}\hat{s}_i^{(\mathrm{n},x)}\hat{s}_K^{(\mathrm{n},N_{\mathrm{x}})}\vec{c}_i^{(\mathrm{n},x)}\vec{e}_i^T & Q(\hat{r})
\end{bmatrix}.
\end{aligned}
$}
\endgroup
\end{equation}

\section{Mathematical Discussion on Adapted Incomplete Partial Fraction CRAM}
\label{ap:aipfcram}
This adapted CRAM relies on decomposing the system provided in Equation \ref{eq:augmentgov} back into its constituents provided in Equations \ref{eq:batemanp}, \ref{eq:ug}, and \ref{eq:cog}.
Following the discussion of the incomplete partial fraction form of CRAM given in \citet{romano2021depletion}, it is efficient to sequentially solve the linear system
\begin{gather}
\label{eq:fullcram}
\left(\mat{A}(\hat{r})t-\theta_\ell\mat{I}_{K(2+KN_{\mathrm{x}})}\right)\vec{y}_\ell=\vec{z}_{\ell-1}
\end{gather}
for each pole $\theta_\ell$ and associated coefficient $\widetilde{\alpha}_\ell$ \citep{maria2016higher}, and use the obtained $\vec{y}_\ell$ to update a cumulative estimate of $\vec{z}(t;\hat{r})$.
After each solution routine,
\begin{gather}
\label{eq:zell}
\vec{z}_{\ell}=\vec{z}_{\ell-1}+2\operatorname{Re}\left(\widetilde{\alpha}_\ell\vec{y}_\ell\right)
\end{gather}
and the approach terminates on iteration $\ell=L$ such that $\vec{z}(t;\hat{r})\approx\alpha_0\vec{z}_L$.
In these expressions, $\vec{z}_0$ maintains its definition given in Equation \ref{eq:augmentic} and $\mat{I}_{K(2+KN_{\mathrm{x}})}\in\mathbb{R}^{K(2+KN_{\mathrm{x}})\times K(2+KN_{\mathrm{x}})}$ is an identity matrix.
The adapted CRAM exploits the structure of $\mat{A}(\hat{r})$ to avoid the costly full-system solve by first rewriting Equation \ref{eq:fullcram} as
\begin{gather}
\left(\mat{A}(\hat{r})t-\theta_\ell\mat{I}_{K(2+KN_{\mathrm{x}})}\right)
\begin{bmatrix}
\vec{y}^{(0)}_\ell \\
\left[\vec{y}_{i,\ell}^{(\mathrm{n},x)}\right]_{\substack{i=1,\ldots,K\\x}} \\
\vec{y}^{(2)}_\ell
\end{bmatrix}
=
\begin{bmatrix}
\bm{n}^{(0)}_{\ell-1} \\
\left[\vec{u}_{i,\ell-1}^{(\mathrm{n},x)}\right]_{\substack{i=1,\ldots,K\\x}} \\
\delta\bm{n}^{(2)}_{\ell-1}
\end{bmatrix}.
\end{gather}
Here, $\vec{y}^{(0)}_\ell$, $\vec{y}_{i,\ell}^{(\mathrm{n},x)}$, and $\vec{y}^{(2)}_\ell$ are each defined to update the cumulative estimates of $\bm{n}^{(0)}(t;\hat{r})$, $\vec{u}_i^{(\mathrm{n},x)}(t;\hat{r})$, and $\delta\bm{n}^{(2)}(t;\hat{r})$, respectively.
Then, $\bm{n}^{(0)}_\ell$, $\vec{u}_{i,\ell}^{(\mathrm{n},x)}$, and $\delta\bm{n}^{(2)}_\ell$ correspond to the segmented components of $\vec{z}_\ell$.
For compactness, the dependence of these quantities on $t$ and $\hat{r}$ is suppressed within the iterations.
Then, defining $\mat{S}_\ell\equiv tQ(\hat{r})-\theta_\ell\mat{I}_K$, the structure of $\mat{A}(\hat{r})$ allows for the definition of three sets of linear systems, each with the same $\mat{S}_\ell$ coefficient matrix,
\begin{align}
\mat{S}_\ell\vec{y}_\ell^{(0)} &= \bm{n}^{(0)}_{\ell-1}, \\
\mat{S}_\ell\vec{y}_{i,\ell}^{(\mathrm{n},x)} &= \vec{u}_{i,\ell-1}^{(\mathrm{n},x)}-t\vec{c}_i^{(\mathrm{n},x)}\vec{e}_i^T\vec{y}_\ell^{(0)}, \\
\mat{S}_\ell\vec{y}_\ell^{(2)} &= \delta\bm{n}^{(2)}_{\ell-1}+t\sum_{j=1}^K\sum_y\sum_{i=1}^K\sum_x\hat{\rho}_{ix,jy}\hat{s}_i^{(\mathrm{n},x)}\hat{s}_j^{(\mathrm{n},y)}\vec{c}_i^{(\mathrm{n},x)}\vec{e}_i^T\vec{y}_{j,\ell}^{(\mathrm{n},y)}. \label{eq:rhs2ns}
\end{align}
Note that the solution to the previous set of systems contributes to the right-hand-side shift of the next set.
Using dense matrix scaling models, this structure enables the performance of a single $\mathcal{O}(K^3)$ LU factorization and then $2+KN_{\mathrm{x}}$ subsequent $\mathcal{O}(K^2)$ substitution routines for each $\ell$ iteration.

After these systems are solved, equivalent to Equation \ref{eq:zell}, updates are performed with
\begin{gather}
\bm{n}_\ell^{(0)}=\bm{n}^{(0)}_{\ell-1}+2\operatorname{Re}\left(\widetilde{\alpha}_\ell\vec{y}_\ell^{(0)}\right), \\
\vec{u}_{i,\ell}^{(\mathrm{n},x)}=\vec{u}_{i,\ell-1}^{(\mathrm{n},x)}+2\operatorname{Re}\left(\widetilde{\alpha}_\ell\vec{y}_{i,\ell}^{(\mathrm{n},x)}\right), \\
\delta\bm{n}^{(2)}_\ell=\delta\bm{n}^{(2)}_{\ell-1}+2\operatorname{Re}\left(\widetilde{\alpha}_\ell\vec{y}_\ell^{(2)}\right)
\end{gather}
and the iterations are initiated with
\begin{gather}
\bm{n}^{(0)}_0=\bm{n}_0, \\
\vec{u}_{i,0}^{(\mathrm{n},x)}=\vec{0}, \\
\delta\bm{n}^{(2)}_0=\vec{0}.
\end{gather}
Upon termination,
\begin{gather}
\bm{n}^{(2)}(t;\hat{r})\approx\alpha_0\bm{n}^{(0)}_L+\alpha_0\delta\bm{n}^{(2)}_L.
\end{gather}

\section{Toy-Problem Correlation Matrix}
\label{app:toy_correlation_matrix}

The correlation matrix used for the eight stochastic reaction channels was constructed synthetically. It is not derived from nuclear-data covariance files; rather, it is introduced to test the second-order formalism in the presence of nonzero correlations between uncertain reaction channels.

Two dimensionless weight vectors were defined,
\begin{equation}
  \vec{w}^{\,F}
  =
  \begin{bmatrix}
  0.65 & 0.60 & 0.60 & 0.55 & 0.50 & 0.45 & 0.50 & 0.40
  \end{bmatrix}^{T},
  \label{eq:toy_wF}
\end{equation}
and
\begin{equation}
  \vec{w}^{\,S}
  =
  \begin{bmatrix}
  0.15 & 0.10 & 0.10 & 0.20 & 0.15 & 0.10 & 0.15 & 0.05
  \end{bmatrix}^{T}.
  \label{eq:toy_wS}
\end{equation}
These vectors can be interpreted as the effect on each stochastic channel of two common latent sources of uncertainty. For two distinct channels $X_\alpha$ and $X_\beta$, the off-diagonal correlation coefficient was defined as
\begin{equation}
  \rho_{\alpha\beta}
  =
  w^{F}_{\alpha}w^{F}_{\beta}
  +
  w^{S}_{\alpha}w^{S}_{\beta},
  \qquad
  \alpha\neq\beta.
  \label{eq:toy_corr_offdiag}
\end{equation}
The diagonal entries were then set to unity,
\begin{equation}
  \rho_{\alpha\alpha}=1,
  \label{eq:toy_corr_diag}
\end{equation}
as required for a correlation matrix.

\begin{figure}[!htbp]
  \centering
  \safeincludegraphics[width=0.70\textwidth]{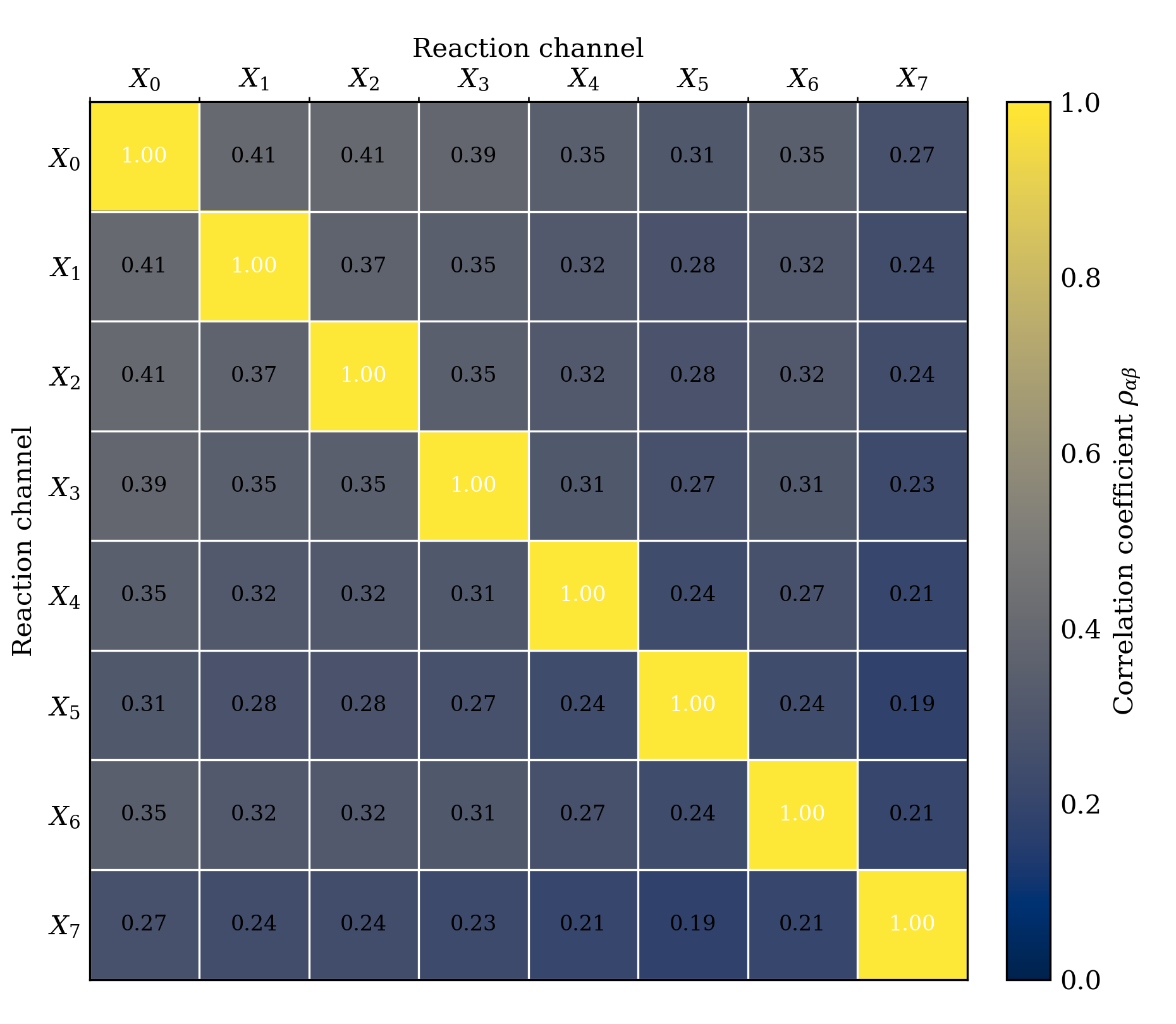}
  \caption{Correlation matrix $\mat{\rho}$ used for the stochastic reaction
  channels in the ten-nuclide toy problem.}
  \label{fig:toy_correlation_matrix}
\end{figure}

This construction is equivalent to assigning each standardized perturbation a shared component, controlled by the two common factors, plus an independent residual component that brings the total variance of each channel to one.
The resulting matrix is shown in Figure~\ref{fig:toy_correlation_matrix}.

\FloatBarrier

\section{Additional Reaction-Rate Variance Results}
\label{app:tp2_additional_variances}

\begin{figure*}[!htbp]
    \centering
    \safeincludegraphics[width=0.78\textwidth]{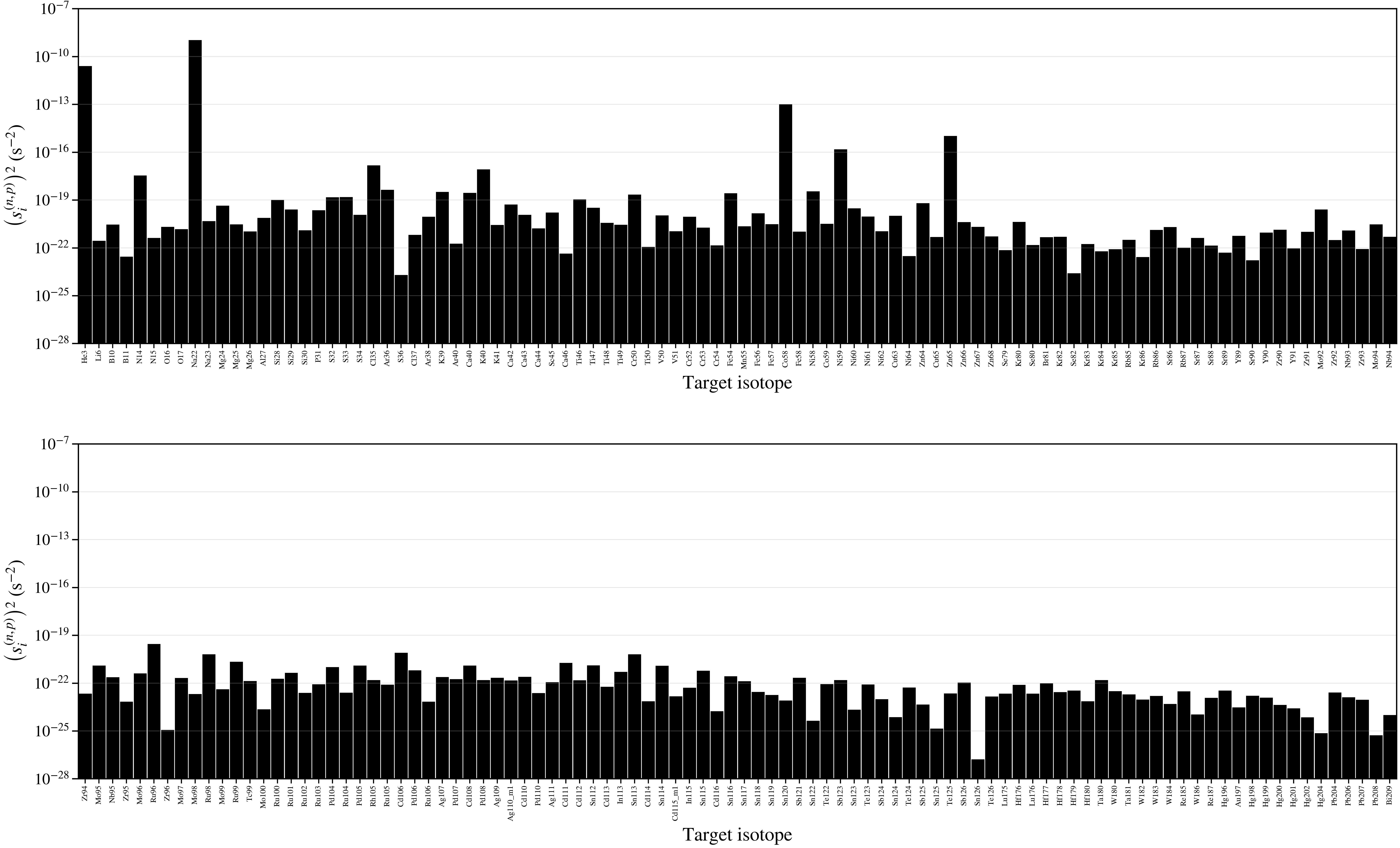}
    \caption{Variance of the $(n,p)$ reaction-rate channels.}
    \label{fig:tp2_np_variance}
\end{figure*}

\begin{figure*}[!htbp]
    \centering
    \safeincludegraphics[width=0.78\textwidth]{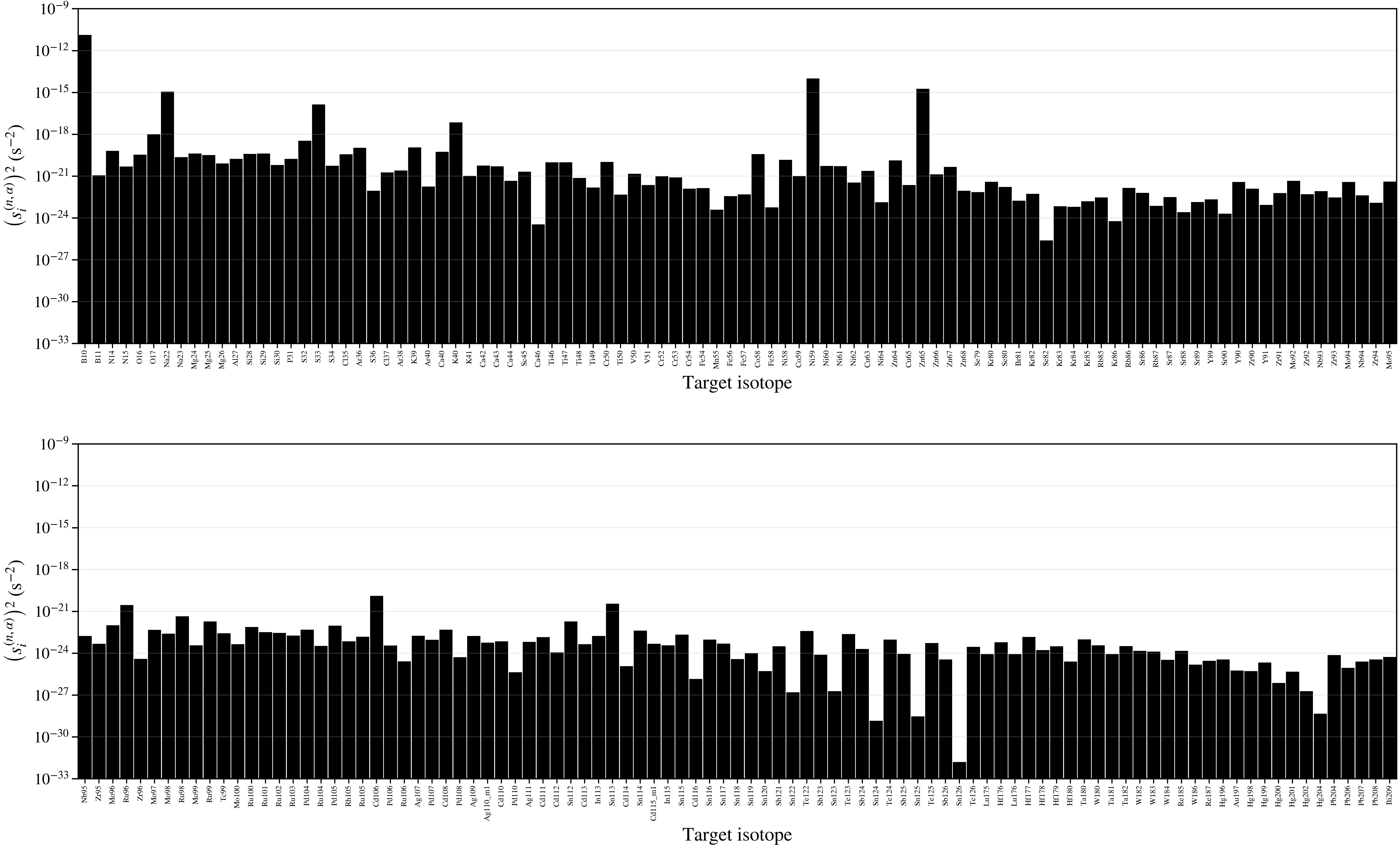}
    \caption{Variance of the $(n,\alpha)$ reaction-rate channels.}
    \label{fig:tp2_na_variance}
\end{figure*}

\begin{figure*}[!htbp]
    \centering
    \safeincludegraphics[width=0.78\textwidth]{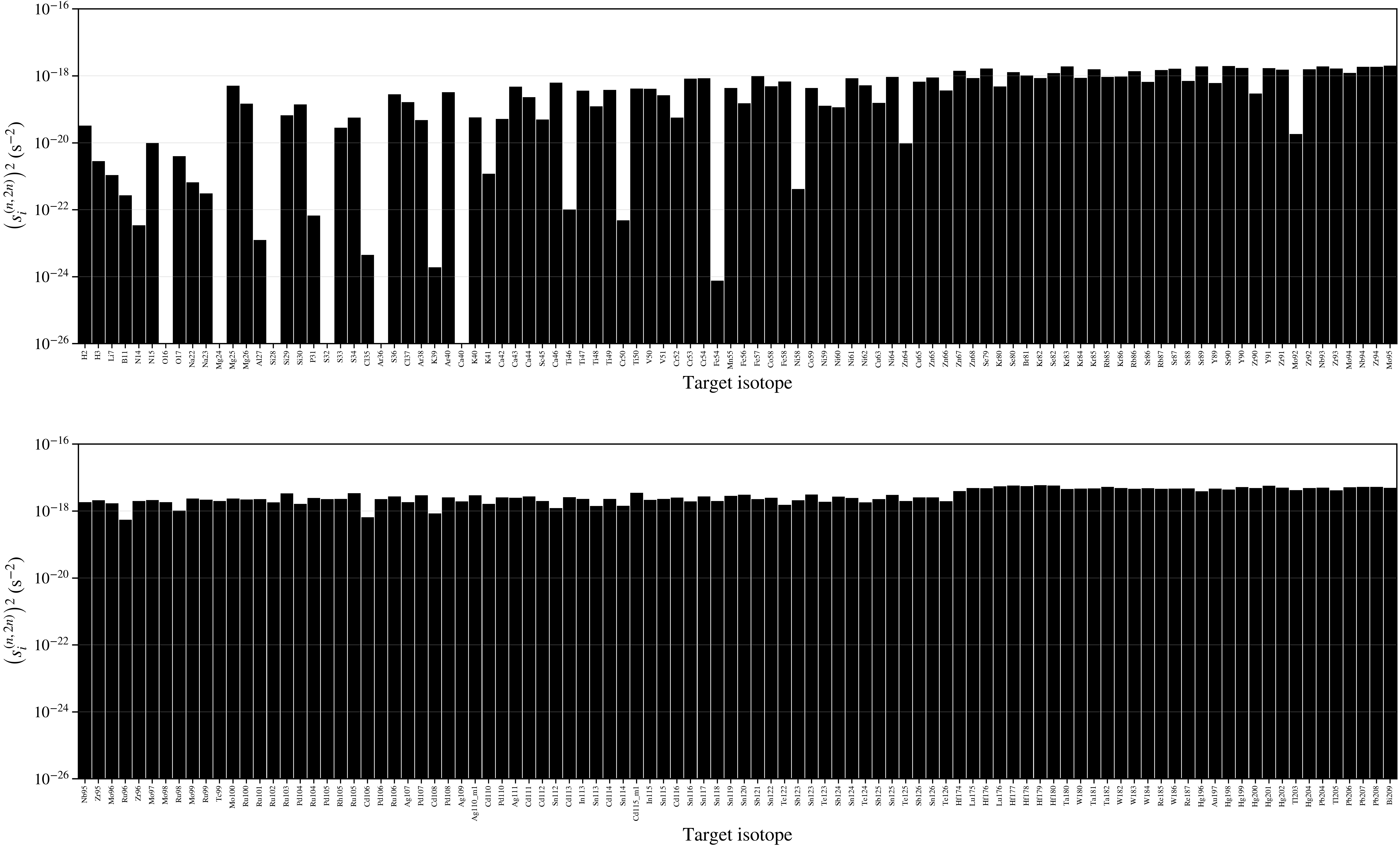}
    \caption{Variance of the $(n,2n)$ reaction-rate channels.}
    \label{fig:tp2_n2n_variance}
\end{figure*}

\begin{figure*}[!htbp]
    \centering
    \safeincludegraphics[width=0.78\textwidth]{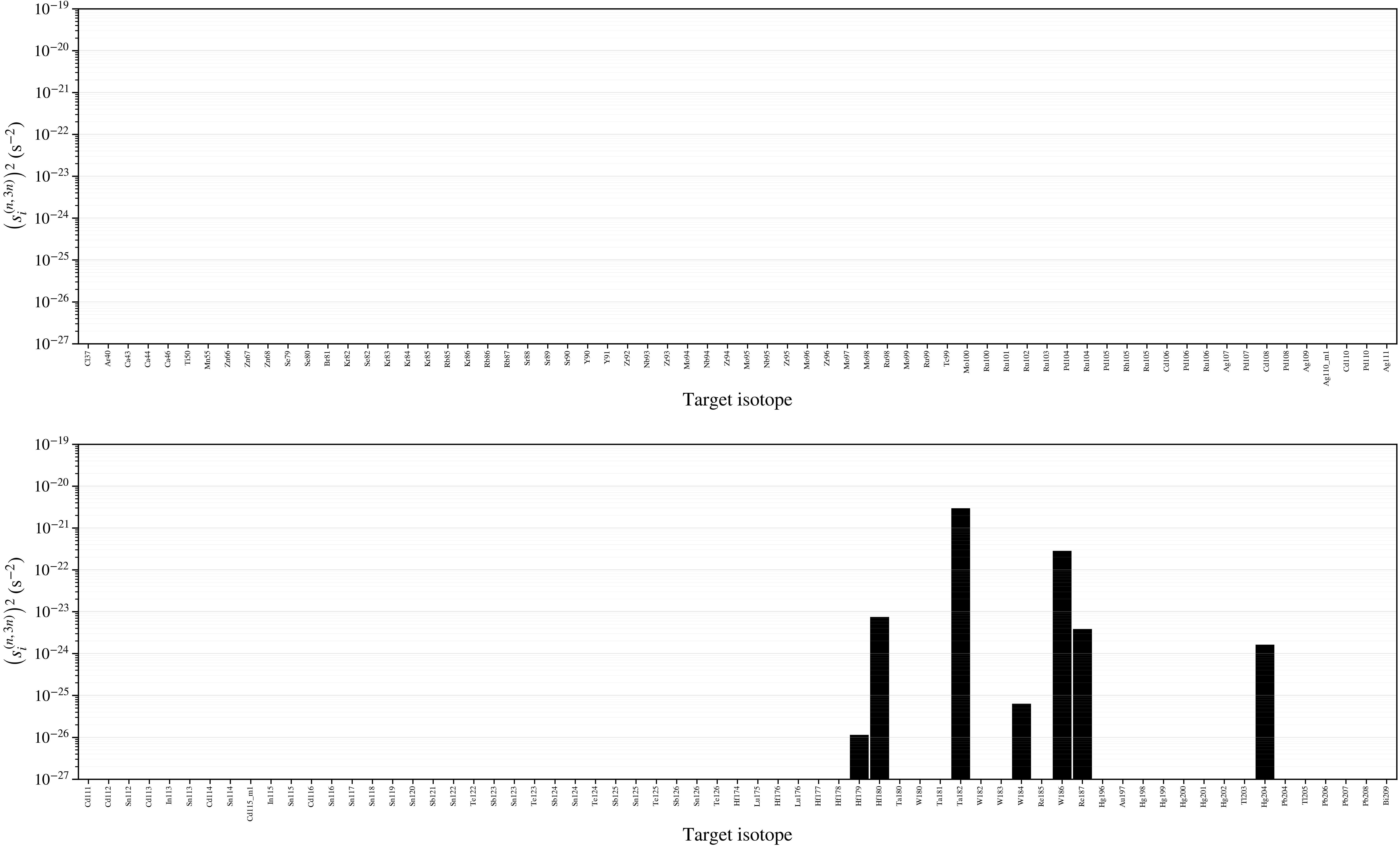}
    \caption{Variance of the $(n,3n)$ reaction-rate channels.}
    \label{fig:tp2_n3n_variance}
\end{figure*}

Figures~\ref{fig:tp2_np_variance}--\ref{fig:tp2_n3n_variance} show the nuclide-specific reaction-rate variances for the remaining active reaction families in the ITER port-plug application.

\FloatBarrier

\section{Representative Reaction-Rate Correlations}
\label{app:tp2_pairplot}

\begin{figure*}[!htbp]
    \centering
    \safeincludegraphics[width=0.88\textwidth]{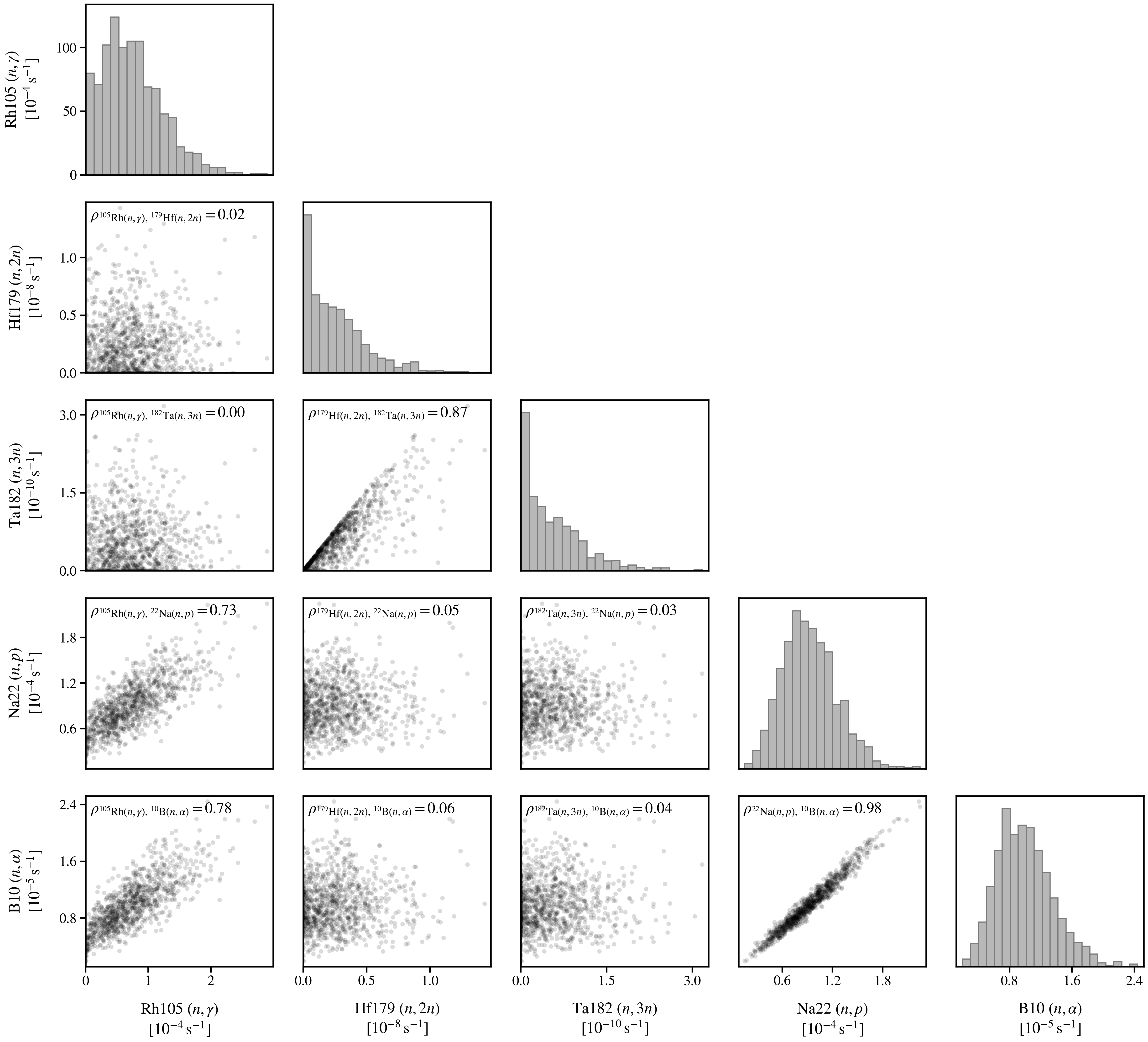}
    \caption{Representative pairwise distributions of selected reaction-rate channels.}
    \label{fig:tp2_pairplot}
\end{figure*}

Figure~\ref{fig:tp2_pairplot} shows representative pairwise distributions for selected reaction-rate channels in the ITER port-plug application.

\FloatBarrier

\end{document}